\documentclass[aps,prd,preprint,a4paper,showpacs,nofootinbib,superscriptaddress]{revtex4-2}
\usepackage{bm}
\usepackage{indentfirst}
\usepackage{amsmath}
\usepackage{graphicx}
\usepackage{amsmath}
\usepackage{CJK}

\usepackage{amssymb}
\usepackage{subfigure}
\usepackage{amssymb}
\usepackage{hyperref}
\usepackage{epstopdf}
\usepackage[section]{placeins}

\usepackage[utf8]{inputenc}
\hypersetup{
    colorlinks=true,
    linkcolor=red,
    citecolor=blue,
}
\usepackage{color}
\usepackage[T1]{fontenc}
\usepackage{txfonts}
\usepackage{orcidlink}

\begin{document}

\title{Rotating scalarized supermassive black holes}

\author{Shoupan Liu}
\email{shoupan_liu@163.com}
\address{\textit{Center for Gravitation and Cosmology, College of Physical Science and Technology, Yangzhou University, Yangzhou 225009, China}}

\author{Yunqi Liu}
\email{yunqiliu@yzu.edu.cn (corresponding author)}
\address{\textit{Center for Gravitation and Cosmology, College of Physical Science and Technology, Yangzhou University, Yangzhou 225009, China}}

\author{Yan Peng}
\email{yanpengphy@163.com}
\address{\textit{School of Mathematical Sciences, Qufu Normal University, Qufu, Shandong 273165, China}}

\author{Cheng-Yong Zhang}
\email{zhangcy@email.jnu.edu.cn}
\address{\textit{Department of Physics and Siyuan Laboratory, Jinan University, Guangzhou 510632, China}}

\baselineskip=0.5 cm

\begin{abstract}
In this study, we investigate rotating black hole solutions within a scalar–Gauss–Bonnet gravity framework that incorporates a squared Gauss–Bonnet term. By employing a quadratic–exponential coupling function between the scalar field and the Gauss–Bonnet invariant, we derive both the standard General Relativity solutions and novel scalarized black hole configurations. Utilizing a pseudo-spectral method to solve the coupled field equations, we examine how black hole spin and coupling constants influence the existence and properties of these solutions. Our findings reveal that both the rotation of the black hole and the squared coupling term effectively constrain the parameter space available for scalarization. Moreover, we demonstrate that, over a wide range of parameters, scalarized black holes exhibit higher entropy than Kerr black holes of equivalent mass and spin, indicating that they are thermodynamically favored. 
These results significantly expand the phase space of black holes in modified gravity theories.

\end{abstract}

\maketitle

\newpage

%%%%%%%%%%%%%%%%%%%%%%%%%%%
%%%%%%%%%%%%%%%%%%%%%%%%%%%
\section{Introduction}
\label{Introduction}
%%%%%%%%%%%%%%%%%%%%%%%%%%%
%%%%%%%%%%%%%%%%%%%%%%%%%%%

In general relativity (GR), stationary vacuum black holes (BHs) are uniquely described by the Kerr metric~\cite{chrusciel2012stationary} and are therefore fully characterized by their mass, electric charge, and angular momentum \cite{kerr2009kerr}.
This remarkable simplicity is encapsulated in the “no-hair” conjecture \cite{PhysRevLett.34.905,PhysRevLett.26.331,doi:10.1142/S0218271815420146} and the Kerr hypothesis\cite{Herdeiro2023}, both of which have long been cornerstones of GR.
Moreover, the no-hair conjecture has been extended to alternative theories, such as Brans-Dicke theories, certain classes of scalar-tensor theories, and Gallilen models of gravity \cite{Pani:2011gy, Herdeiro:2014goa, Babichev:2013cya}.

Over the past decades, however, a growing body of work in modified gravity theories has revealed the existence of “hairy” BH solutions that carry additional charges beyond the traditional parameters\cite{Cadoni:2009xm, Cardoso:2013fwa, Kleihaus:2015iea, Silva:2018qhn}.
Notable examples include black holes in the presence of Yang-Mills fields~\cite{Volkov:1989fi, Bizon:1990sr,Greene:1992fw,Maeda:1993ap}, Skyrme~\cite{Luckock:1986tr,Droz:1991cx}, conformally coupled scalar fields~\cite{Bekenstein:1974sf}, and dilatonic or colored BHs within Einstein-dilaton-Gauss-Bonnet theory~\cite{Torii:1996yi,Kanti:1996gs,Kleihaus:2015aje,Kleihaus:2011tg,Guo:2008hf}.
Other cases involve rotating~\cite{Ayzenberg:2014aka, Maeda:2009uy,Ohta:2010ae} or shift-symmetric Galileon hairy BHs~\cite{Sotiriou:2014pfa,Benkel:2016rlz}.
Recent studies have provided significant insights into the phase diagram of gravitational systems beyond GR~\cite{Antoniou:2021zoy, Liu:2022eri,Jiang:2023yyn,Zhang:2022cmu,Doneva:2022yqu,Doneva:2022ewd,Doneva:2023kkz,Liu:2022fxy,Zhang:2023jei,Lai:2023gwe,Jiang:2023yyn,Minamitsuji:2023uyb}. 
This is particularly interesting in theories that may be affected by tachyonic instability~\cite{Damour:1993hw, Cardoso:2013opa, Cardoso:2013fwa, Zhang:2014kna,Liu:2022eri}, which can trigger {\it spontaneous scalarization}.
This mechanism often appears in models where a real scalar field is non-minimally coupled to specific source terms.
The presence of non-minimal coupling acts as an effective mass squared term in the scalar field's equation of motion; when this effective mass squared term becomes negative, a tachyonic instability may ensue.
Such source terms could include geometrically invariant quantities, such as the Gauss-Bonnet invariant in extended scalar-Gauss-Bonnet (sGB) theory~\cite{Doneva:2017bvd, Silva:2017uqg, Antoniou:2017acq, Cunha:2019dwb, Dima:2020yac, Herdeiro:2020wei, Berti:2020kgk, Corelli:2022pio, Corelli:2022phw}, the Ricci scalar for non-conformally invariant BHs~\cite{Herdeiro:2019yjy}, the Chern-Simons invariant~\cite{Brihaye:2018bgc}, or the Maxwell invariant $F_{\mu\nu}F^{\mu\nu}$~\cite{Herdeiro:2018wub,Zhang:2021edm}.
Depending on the form of the coupling function, the onset of scalarization may follow a linear or a nonlinear route, 
the latter is often referred to as nonlinear scalarization~\cite{Doneva:2021tvn,Blazquez-Salcedo:2022omw,Zhang:2021nnn,Liu:2022fxy,Zhang:2024spn,Gonzalez:2024ifp,Jiang:2023yyn}.
%This nonlinear behavior typically occurs in theories with higher-order coupling functions, where both the coupling function and its first derivative with respect to the scalar field vanish at a stationary point where its second derivative does not.

Generally, scalarization is unattainable for BHs with intermediate and supermassive mass scales~\cite{Antoniou:2021zoy, Liu:2022eri,Jiang:2023yyn,Zhang:2022cmu,Doneva:2022yqu,Doneva:2022ewd}. Consquently, the prevailing expectation is that BHs in these mass ranges should be well described by the Kerr metric, even when solar-mass BHs are not.
However, Ref.~\cite{eichhorn2023breaking} considers a model in which the Gauss-Bonnet term is quadratically coupled to a scalar field. 
By studying the model in spherical symmetry, they found that unlike in the conventional framework, BHs can undergo spontaneous scalarization within a finite mass window, which may include supermassive BHs.
Such deviations are particularly interesting because they offer a potential observation window through future gravitational wave detectors like LISA, Taiji, and TianQin to probe into the underlying gravitational theory \cite{Danzmann:1997hm,Hu:2017mde,TianQin:2015yph,Li:2024rnk}.
Building on the preliminary rotating solution analysis in Appendix A of Ref.~\cite{eichhorn2023breaking}, we present the first systematic parameter study of stationary axisymmetric scalarized black holes using pseudo-spectral methods. Our work quantifies scalarization suppression by spin and coupling constants, establishes entropy dominance over Kerr solutions, and maps the full existence domain beyond spherical symmetry.

In the present paper, we extend these investigations to rotating BHs.
By employing a quadratic-exponential coupling function, we explore how the inclusion of a squared Gauss-Bonnet term modifies the spectrum of rotating BH solutions. 
Our study addresses two central questions: How does the presence of angular momentum affect the onset and domain of scalarization?
And how do squared curvature corrections influence the stability and thermodynamic properties of these scalarized configurations compared to their Kerr counterparts?
To answer these questions, we numerically solve the full set of modified field equations under stationary and axisymmetric assumptions using the Chebyshev pseudo-spectral method combined with a Newton-Raphson iterative scheme.
Our analysis reveals that both the spin parameter and the additional squared coupling act to suppress the scalarization, thereby reducing the parameter space in which scalarized solutions exist. 
Nevertheless, in many regimes the scalarized BHs exhibit higher entropy than Kerr BHs of the same mass and spin, suggesting that they are thermodynamically favored.

The paper is organized as follows. Sec.\ref{Theoretical setup} establishes the theoretical framework including action, field equations, and numerical boundary conditions. 
We further analyze the stability of general relativistic black holes under scalar perturbations and discuss the mechanisms of sGB coupling that drive spontaneous. Sec.\ref{Numerical method} details the numerical methodology employed to solve the system, focusing on the Chebyshev pseudo-spectral approach and the treatment of boundary conditions in compactified radial coordinates. Numerical results are presented in Sec.\ref{Numerical Results}, where we systematically explore the parameter space to characterize scalarized black hole solutions and their dependence on spin, coupling constants. Finally, Sec.\ref{Conclusions} summarizes our conclusions. Throughout this work, we adopt units $G$ = $c$ = 1.

%%%%%%%%%%%%%%%%%%%%%%%%%%%
%%%%%%%%%%%%%%%%%%%%%%%%%%%
\section{Theoretical setup}
\label{Theoretical setup}
%%%%%%%%%%%%%%%%%%%%%%%%%%%
%%%%%%%%%%%%%%%%%%%%%%%%%%%
\subsection{Action and basic equations}

We consider a four-dimensional sGB gravity theory with squared curvature corrections, defined by the action \cite{eichhorn2023breaking}:
\begin{eqnarray}\label{Act}
    S=\frac{1}{16\pi} \int d^4x \sqrt{-g} [R-(\partial \phi)^2 + \alpha_1 F(\phi) \mathcal{G} - 2\alpha^3_2 F(\phi)(\psi \mathcal{G} - \frac{\psi ^2}{2})],
\end{eqnarray}
where the Gauss-Bonnet invariant is
\begin{eqnarray}\label{GB term}
   \mathcal{G}=R^2-4R_{\alpha \beta}R^{\alpha \beta}+R_{\alpha \beta \gamma \sigma}R^{\alpha \beta \gamma \sigma},
\end{eqnarray}
with $R$, $R_{\alpha \beta}$ and $R_{\alpha \beta \gamma \sigma}$ denoting the Ricci scalar, Ricci tensor, and Riemann tensor respectively.
The theory contains two dimensionless coupling constants $\alpha_1$ and $\alpha_2$ (with dimensions of length squared), a dimensionless real scalar field $\phi$, and an auxiliary field $\psi$ with dimensions inverse length to the fourth, which is the same as $\mathcal{G}$.

Varying the action with respect to the metric yields the Einstein field equations:
\begin{eqnarray}\label{EOM}
    \mathcal{E_{\mu \nu}} = G_{\mu \nu} - T_{\mu \nu} = 0,
\end{eqnarray}
where the effective energy-momentum tensor contains contributions from both scalar fields and curvature couplings:
\begin{eqnarray}\label{EM}
    T_{\mu \nu} = \nabla_{\mu} \phi \nabla_{\nu} \phi - \frac{1}{2}g_{\mu \nu}\left[(\nabla \phi)^2 + \alpha^3_2 \psi^2 F(\phi) \right] + 4 P_{\mu \alpha \nu \beta} 
    \nabla^{\alpha} \nabla^{\beta} \left[(\alpha_1 - 2\alpha^3_2 \psi) F(\phi)\right],
\end{eqnarray}
with the double dual Riemann tensor
\begin{eqnarray}\label{P}
   P_{\mu \nu \alpha \beta} \equiv \frac{1}{4} \epsilon_{\mu \nu \gamma \delta} R^{\rho \sigma \gamma \delta} \epsilon_{\rho \sigma \alpha \beta} 
   = R_{\mu \nu \alpha \beta} + 2g_{\mu [\beta}R_{\alpha]\nu} + 2g_{\nu[\alpha}R_{\beta]\mu} + R g_{\mu[\alpha}g_{\beta]\nu}.
\end{eqnarray}

The scalar field equations follow from variation with respect to $\phi$:
\begin{eqnarray}\label{KG}
    \square \phi + \left [\alpha_1 \mathcal{G} - 2 \alpha^3_2 \left (\psi \mathcal{G} - \frac{\psi^2}{2} \right )\right ] \frac{F'(\phi)}{2} = 0,
\end{eqnarray}
where the prime $``'"$ denotes the derivative with respect to scalar field $\phi$.
The scalar $\psi$ acts as a Lagrange multiplier, which is not dynamical and satisfies the following equation
\begin{eqnarray}\label{LM}
    \psi - \mathcal{G} = 0.
\end{eqnarray}
From Eq.\eqref{LM}, one can see that actually the action \eqref{Act} includes a term of $\mathcal{G}^2$.

In this paper, we will consider a quadratic-exponential coupling $F(\phi) = \frac{1}{\kappa}(1 - e^{-\kappa \phi^2})$ to study the properties of scalarized solutions where $\kappa$ is a constant.
The coupling function has a stationary point that satisfies $F(\phi)=0,~F'(\phi)=0$.
By the properties of the coupling function, it is easy to find that the equations of motion, Eqs.(\ref{EOM}) and (\ref{KG}), allow the GR solutions.
For convenience, we set $\phi=0$ at the stationary point.
However, the GR configuration may be unstable for some model parameters.
To show the instability of GR BHs and the potential scalarization, we turn to the linear perturbation analysis.
By linearizing Eq.(\ref{KG}) around $\phi=0$, we obtain the equation of motion governing the scalar perturbation $\phi_p$ as follows
\begin{eqnarray}\label{perturbation Eq}
   \square \phi_p =m^2_{eff} \phi_p, ~~~~~m^2_{eff}=-\frac{1}{2} \left( \alpha_1 \mathcal{G}-\alpha_2^3 \mathcal{G}^2 \right)  F''(0) 
\end{eqnarray}
where we have substituted the equation of motion Eq.(\ref{LM}) into Eq.(\ref{KG}), $m^2_{eff}$ is the effective mass squared of the perturbation.
The different behavior of the derivative $F''(0)$ leads to a different scalarization mechanism.
For the quadratic-exponential coupling in this work, we have the value $F''(0)\neq0$ which means the squared effective mass would negative enough that make the perturbations tachyonically unstable \cite{PhysRevLett.125.231101,PhysRevD.102.084060,PhysRevD.102.124056,PhysRevD.102.104027,PhysRevLett.120.131104,Liu:2022eri},
as a result the GR BH solution would be unstable and undergo spontaneous scalarization.

\subsection{Equations of motion and boundary behaviors }
To construct stationary axisymmetric solutions, we adopt a quasi-isotropic coordinate system with metric Ansatz \cite{fernandes2023new} :
\begin{eqnarray}\label{LineElement}
    ds^2=-f \mathcal{N}^2 dt^2 + \frac{g}{f} \left [h \left (dr^2 + r^2 d \theta ^2 \right ) + r^2 sin^2 \theta \left (d \varphi - \frac{W}{r}(1 - \mathcal{N})dt
     \right )^2 \right ],
\end{eqnarray}
where $\mathcal{N}  = 1 - r_H / r$ regularizes the horizon at $r=r_H$, and the dimensionless functions
$f$, $g$, $h$, and $W$ depend on the radial $r$ and the angular parameter $\theta$. 

To solve the system, we employ the following equations that diagonalize the Einstein tensor with respect to the differential operator $\partial^2_r + r^{-2} \partial^2_{\theta}$,
\begin{subequations}\label{CEinsteinFieldEq}
\begin{align}
   0=&~-\mathcal{E}^{\mu}_{\ \mu} + 2 \mathcal{E}^{t}_{\ t} + \frac{2W r_H}{r^2} \mathcal{E}^{\varphi}_{\ t},\label{CEinsteinFieldEq-1} \\
   0=&~\mathcal{E}^{\varphi}_{\ t},\label{CEinsteinFieldEq-2} \\ 
   0=&~\mathcal{E}^{r}_{\ r} + \mathcal{E}^{\theta}_{\ \theta},\label{CEinsteinFieldEq-4} \\
   0=&~\mathcal{E}^{\varphi}_{\ \varphi} - \frac{W r_H}{r^2} \mathcal{E}^{\ \varphi}_{t} - \mathcal{E}^{r}_{\ r} - \mathcal{E}^{\theta}_{\ \theta} \label{CEinsteinFieldEq-4}
\end{align}
\end{subequations}

Asymptotic flatness imposes boundary conditions at spatial infinity ($r \rightarrow \infty$) :
$\lim\limits_{r \to \infty}{f}=\lim\limits_{r \to \infty}{g}=\lim\limits_{r \to \infty}{h}=1$, 
$\lim\limits_{r \to \infty}{2 r_H r^2 \partial_{r} W + \left( 2 r_H + r^2 \partial_{r} f \right)^2 \chi}=0$ 
with $ \chi \equiv J/M^2$ defining the dimensionless spin parameter, and
$\lim\limits_{r \to \infty} \phi=\lim\limits_{r \to \infty}\psi=0$. 
Axial symmetry and regularity require vanishing angular derivatives at the symmetry axis $\theta=0, \pi$ :
$\partial_{\theta}f=\partial_{\theta}g=\partial_{\theta}h=\partial_{\theta}W=\partial_{\theta}\phi=\partial_{\theta}\psi=0$. 
Furthermore, the absence of conical singularities imposes that on the symmetry axis: $h=1$ for $\theta=0, \pi$. The event horizon boundary conditions are 
$f-r_H \partial_{r}f \big|_{r=r_H} = g + r_H \partial_{r}g \big|_{r=r_H} = \partial_{r}h \big|_{r=r_H} = W + r_H \partial_{r}W/2\big|_{r=r_H}=0.$

Most of the physical quantities of interest are encapsulated in the metric functions evaluated either at the horizon or at infinity. 
Starting with the asymptotic quantities, the Arnowitt-Deser-Misner (ADM) mass $M$ and the angular momentum $J$ can be extracted from the asymptotic expansion:
\begin{align}
    & g_{tt}= -f \mathcal{N}^2 + \frac{g \left( 1-\mathcal{N} \right)^2 W^2}{f} \sin^2{\theta} = -1 + \frac{2M}{r} + \mathcal{O} (r^{-2}),\notag\\
    & g_{t \varphi} = -\frac{g r \left( 1-\mathcal{N} \right) W}{f} \sin^2{\theta} = - \frac{2J}{r} \sin^2 \theta  + \mathcal{O} (r^{-2}).
\end{align}
Scalar charge $Q_s$ appears in the far-field behavior:
\begin{eqnarray}\label{scalar charge}
   \phi = \frac{Q_s}{r}+ \mathcal{O}  (r^{-2}).
\end{eqnarray}

For numerical implementation of the Einstein-scalar system via the Chebyshev pseudo-spectral method, we introduce compactified radial coordinate $x=1-2r_H/r$ that map the semi-infinite domain $r\in\left [r_H, \infty \right )$ to a finite interval $x\in\left [-1, 1 \right ]$. This coordinate transformation regularizes the event horizon at $x=-1$ while preserving asymptotic flatness at $x=1$.
Then the BH mass becomes
\begin{eqnarray}\label{ADM}
    M=r_H\left( 1+\partial_{x}f\right)\Big|_{x=1}.
\end{eqnarray}
The horizon geometry is encoded in the induced metric:
\begin{eqnarray}
    d \Sigma^2 = \gamma_{ij} dx^i dx^j = \frac{r_H^2 g}{f} \left[ h d\theta^2 +\sin^2 {\theta}d\varphi^2 \right] \Big|_{x=-1},
\end{eqnarray}
which allows computation of black hole entropy via the Iyer-Wald formalism \cite{wald1993black,iyer1994some}
\begin{align}\label{entropy}
   S = \frac{A_H}{4} + \frac{1}{4} {\int_{\mathcal{H}}^{}  \,d^{2}x} \ \sqrt[]{\gamma} \left (\alpha_1 - 2\alpha^3_2 \psi \right )F(\phi) \widetilde{R},
\end{align}
where $\mathcal{H}$ denotes the horizon, $\gamma$ is the determinant of the induced metric and $\widetilde{R}$ is the Ricci scalar on it. 
The horizon area $A_H$ is explicitly given by
\begin{align}\label{TA}
    & A_H = 2 \pi r^{2}_H\int_{0}^{\pi} \,d\theta \sin \theta \frac{g \enspace \sqrt[ ]{h}}{f}\Bigg | _{x=-1}.
 \end{align}

\section{Numerical method}
\label{Numerical method}

In this section, we give a brief description of the numerical methods.
We implement the Chebyshev pseudo-spectral method in conjunction with the Newton-Raphson method to solve the coupled system governed by Eqs.\eqref{KG}, \eqref{LM} and \eqref{CEinsteinFieldEq}.
This method has been widely employed in studying rotating black hole solutions \cite{fernandes2023new}.
 
In the compactified coordinate, the boundary conditions at horizon $x=-1$ take the following form:
\begin{align}\label{BChorizon}
f - 2 \partial_x f =0,&\quad g + 2 \partial_x g =0,\quad \partial_x h =0,\notag\\
\quad W -  \partial_x W =0, &\quad \partial_{x} \phi =\partial_x \psi =0.
\end{align}
The asymptotic boundary conditions at spatial infinity ($x=1$) can be written as
\begin{align}\label{BCinfinity}
     f = g = h = 1,\quad  \partial_x W + \chi (1 + \partial_x f)^2=0 \quad \text{and} \quad \phi = \psi = 0.
\end{align}

Exploiting equatorial reflection symmetry ($\theta \rightarrow \pi - \theta$), we restrict the angular domain to $\theta \in \left[0,\pi /2 \right]$ and construct spectral expansions for the six functions $\mathcal{F}^{(k)} = \left\{f, g, h,W,\phi,\psi\right\}$ using tensor products of Chebyshev polynomials and cosine basis functions:
\begin{eqnarray}\label{ExpandSerise}
    \mathcal{F}^{(k)}\left(x,\theta\right) = \sum_{i = 0}^{N_x -1} \sum_{j = 0}^{N_{\theta} -1} \alpha^{(k)}_{ij} T_{i}(x) \cos(2j\theta),
\end{eqnarray}
where $T_i(x) = \cos(i \arccos x)$ denotes the $i$-th Chebyshev polynomial, $\alpha^{(k)}_{ij}$ represent spectral coefficients, and $N_x$ and $N_{\theta}$ are the resolutions in the radial and angular directions. In this paper, we mainly use resolutions of $N_x$ = 40 and $N_{\theta}$ = 8 for the computations, with convergence details provided in Appendix.

Substituting the metric Ansatz \eqref{LineElement} into the field equations \eqref{KG},\eqref{LM} and \eqref{CEinsteinFieldEq} generates a system of nonlinear partial differential equations containing the functions and their first and second derivatives $\left(\mathcal{F}^{(k)},\partial_x \mathcal{F}^{(k)},\partial^2_x \mathcal{F}^{(k)},\partial_{\theta} \mathcal{F}^{(k)},\partial^2_{\theta} \mathcal{F}^{(k)},\partial_{x \theta} \mathcal{F}^{(k)}\right)$. Expressing the field equations in the residual form $\mathcal{R}(x,\theta,\partial \mathcal{F}^{(k)})=0$, and substituting the spectral expansions Eq.~\eqref{ExpandSerise} into the residuals, we can calculate the resulting equations at Gauss-Chebyshev points defined by
\begin{eqnarray}
    x_l = \cos\left[\frac{(2l+1)\pi}{2N}\right], \quad \theta_m = \frac{(2m+1)\pi}{4N}, \quad l,m = 0,...,N-1.
\end{eqnarray}
Together with the boundary conditions, there are $N_{\mathcal{F}} \times N_{x} \times N_{\theta}$ algebraic equations for the spectral coefficients $\alpha^{(k)}_{ij}$, which is given by
\begin{eqnarray}\label{Coeffs}
    \alpha^{(k)}_{ij} = \frac{4}{N_x N_{\theta}} \sum_{l = 0}^{N_x -1} \sum_{m = 0}^{N_{\theta} -1} \mathcal{F}^{(k)}(x_l,\theta_m) T_i(x_l) \cos(2j\theta_{m}).
\end{eqnarray}

To solve those algebraic equations with the Newton-Raphson iterative method, one has to provide an initial guess.
Setting BH parameters, we expand the functions of the Kerr solution in the spectral series Eq.\eqref{ExpandSerise} to obtain the spectral coefficients $\alpha^{(k)}_{ij, Kerr}$.
The values of coefficients $\alpha^{(k)}_{ij, Kerr}$ are set as initial guesses to the Newton solver to search for the spectral coefficients $\alpha^{(k)}_{ij}$ of the "potential" scalarized BH solution.

%%%%%%%%%%%%%%%%%%%%%%%%%%%
%%%%%%%%%%%%%%%%%%%%%%%%%%%

%%%%%%%%%%%%%%%%%%%%%%%%%%%
%%%%%%%%%%%%%%%%%%%%%%%%%%%
\section{Numerical results}
\label{Numerical Results}
%%%%%%%%%%%%%%%%%%%%%%%%%%%
%%%%%%%%%%%%%%%%%%%%%%%%%%%
This section presents the numerical investigation of rotating scalarized BH solutions. 
Firstly, setting the parameter $\kappa$ to different values, we show the spectrum of solutions.

%%%%%%%%%%%%%%%%%%%%%%%%%%%
\subsection{The results for different $\kappa$}
\label{}
%%%%%%%%%%%%%%%%%%%%%%%%%%%

\begin{figure}[h]
    \begin{minipage}{0.45\linewidth}
        % \vspace{3pt}
        \centerline{\includegraphics[width=\textwidth]{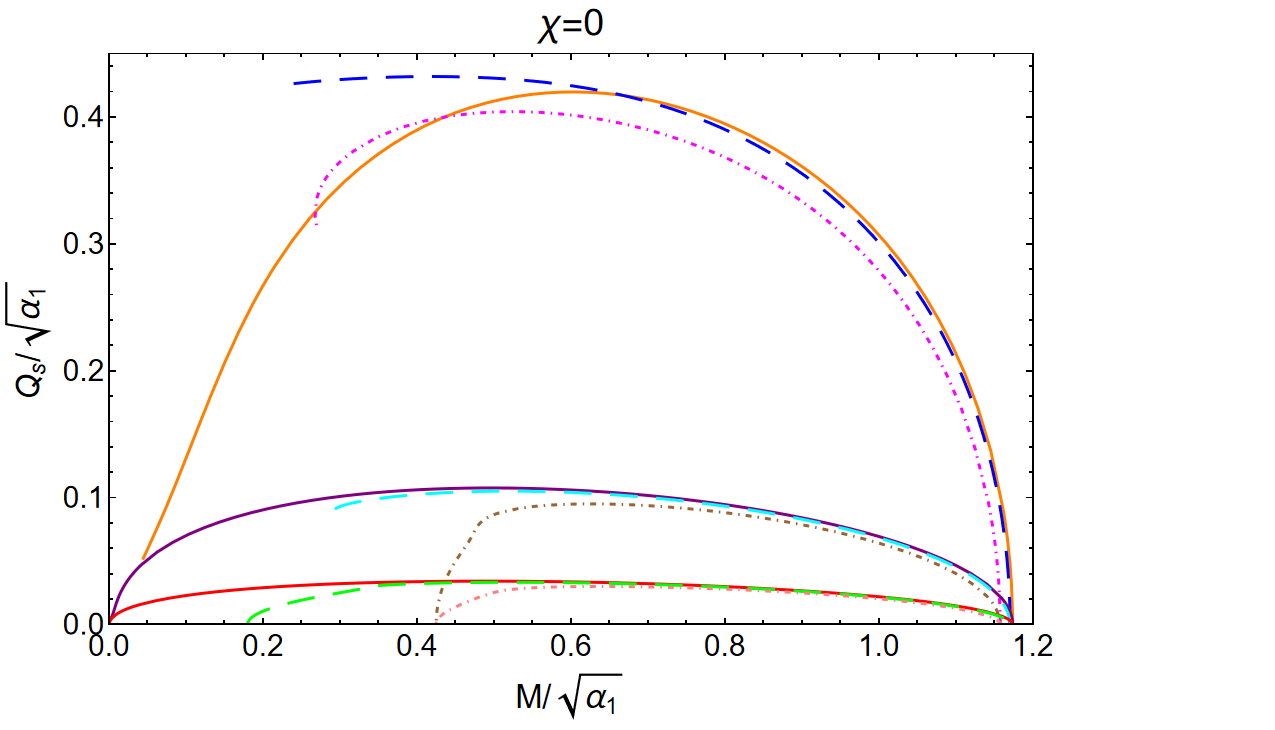}}
        \centerline{\includegraphics[width=\textwidth]{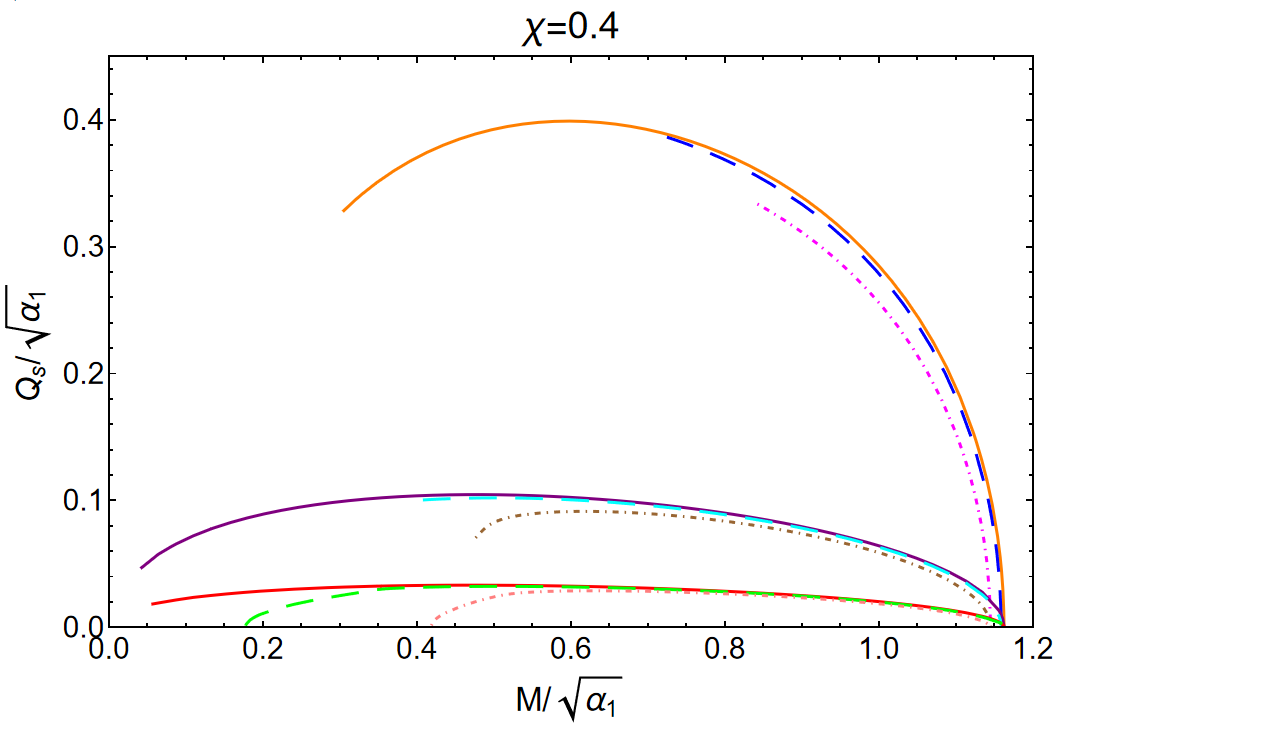}}
    \end{minipage}
   \begin{minipage}{0.45\linewidth}
       % \vspace{3pt}
        \centerline{\includegraphics[width=\textwidth]{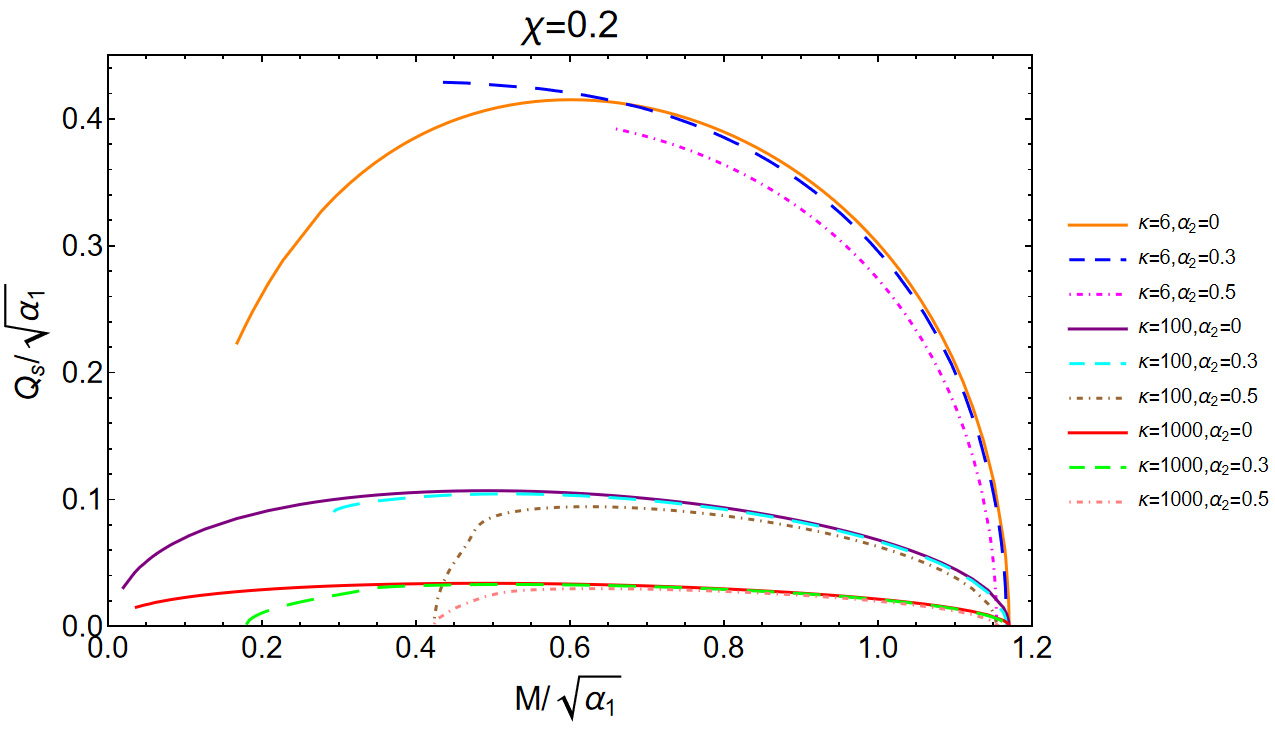}}
        \centerline{\includegraphics[width=\textwidth]{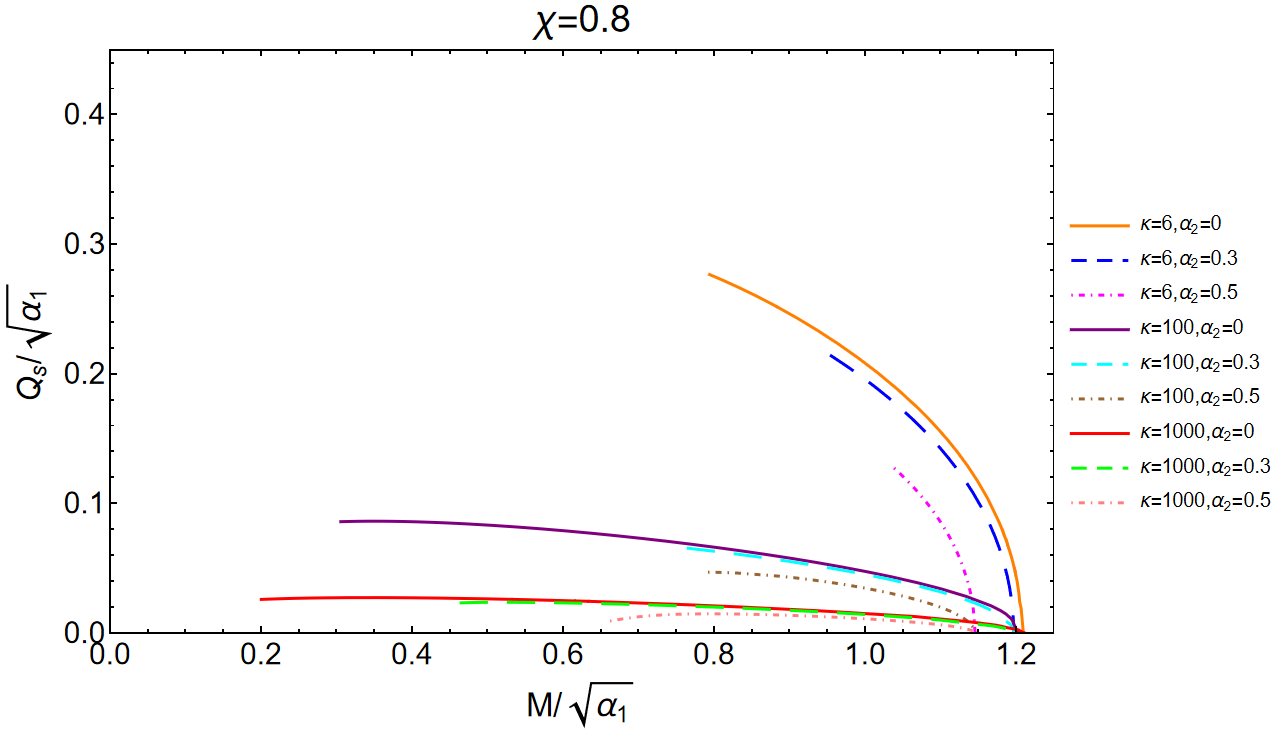}}
   \end{minipage}
    \caption{Scalar charge $Q_s/ \sqrt{\alpha_1}$ as a function of ADM mass $M/ \sqrt{\alpha_1}$ for different parameters.}
    \label{fig:diff kappa}
\end{figure}
Fig. \ref{fig:diff kappa} presents the dimensionless scalar charge $Q_s/ \sqrt{\alpha_1}$ as a function of the ADM mass $M/ \sqrt{\alpha_1}$ for three different values of the parameter $\kappa=6,100,1000$ and three coupling ratios $\alpha_2/\alpha_1=0, 0.3, 0.5$. 
The four panels correspond to increasing values of the dimensionless spin parameter $\chi$ = 0 (non-rotating), $\chi$ = 0.2, 0.4, 0.8.
Within each panel, the solution space is stratified into three distinct coupling regimes: the upper set corresponds to strong exponential coupling ($\kappa=6$), the intermediate set to moderate coupling ($\kappa=100$), and the lower set to weak coupling ($\kappa=1000$). 
This hierarchical structure demonstrates an inverse relation between $\kappa$ and the  scalar charge $Q_s/\sqrt{\alpha_1}$, with the $\kappa$-dependent enhancement becoming more pronounced at higher spins. 
Moreover, for each curve, the left endpoint represents a critical set for the scalarized BH solution, while the right endpoint corresponds to an existence solution which is a marginal BH solution with zero scalar charge.
These endpoints delineate the boundaries of the existence domain of the scalarized BHs, and they will be discussed in detail in the next section.

%critical set and existence solution are the boundaries of the existence domain of the sclarized BHs, and both of them will be discussed in detail in the next section.

The comparative analysis of scalarized BH solutions across different coupling parameters reveals two key interdependent mechanisms governing their existence domains.
First, the constant $\kappa$ plays a decisive role in shaping the scalarization phase space through its inverse relationship with the magnitude of the scalar charge ($Q_s/\sqrt{\alpha_1}$).
This is clearly reflected in the structure of the stratified solution, where the lower $\kappa$ values correspond to upper-branch solutions with enhanced scalar charges. 
In addition, the squared coupling parameter $\alpha_2$ systematically suppresses scalarization across all $\kappa$ regimes.
For fixed $\chi$ and $\kappa$, an increase in $\alpha_2$ reduces both the scalar charge and the parametric range over which viable scalarized configurations exist.
Second, rotational dynamics introduces a spin-dependent suppression mechanism. 
At fixed values of $\kappa$ and $\alpha_2$, increasing the dimensionless spin $\chi$ from $0$ to $0.8$ reduces the existence domain of the solution. 
This spin suppression becomes particularly dominant at higher $\chi$ values.
%, especially in the range $\chi$ > 0.4.
%where Kerr spacetime's ergoregion dynamics begin to significantly influence the scalarization process.
%Third, the hierarchical suppression hierarchy follows $\alpha_2$ > $\chi$ > $\kappa$. The coupling parameter $\alpha_2$ produces the strongest suppression through its direct modification of the additional term included squared Gauss-Bonnet term, effectively stabilizing GR solutions against tachyonic instabilities.
To isolate the $\alpha_2$-$\chi$ interaction while maintaining sufficient scalarization strength for accurate measurement, we focus on case $\kappa=6$ for a further detailed analysis of the properties of the solution in the following section. %This choice ensures scalar charge magnitudes remain above numerical uncertainty thresholds (ΔQ_s/Q_s < 5%) across the full χ ∈ [0,0.8] range.

%%%%%%%%%%%%%%%%%%%%%%%%%%%
\subsection{The results for $\kappa=6$}
\label{k6}
%%%%%%%%%%%%%%%%%%%%%%%%%%%

In this section, we present the domain of existence and the physical properties of rotating scalarized BHs with the coupling function $F(\phi) = (1 - e^{-6 \phi^2})/6$.

\begin{figure}[h!]
    \begin{minipage}{0.4\linewidth}
        % \vspace{3pt}
        \centerline{\includegraphics[width=\textwidth]{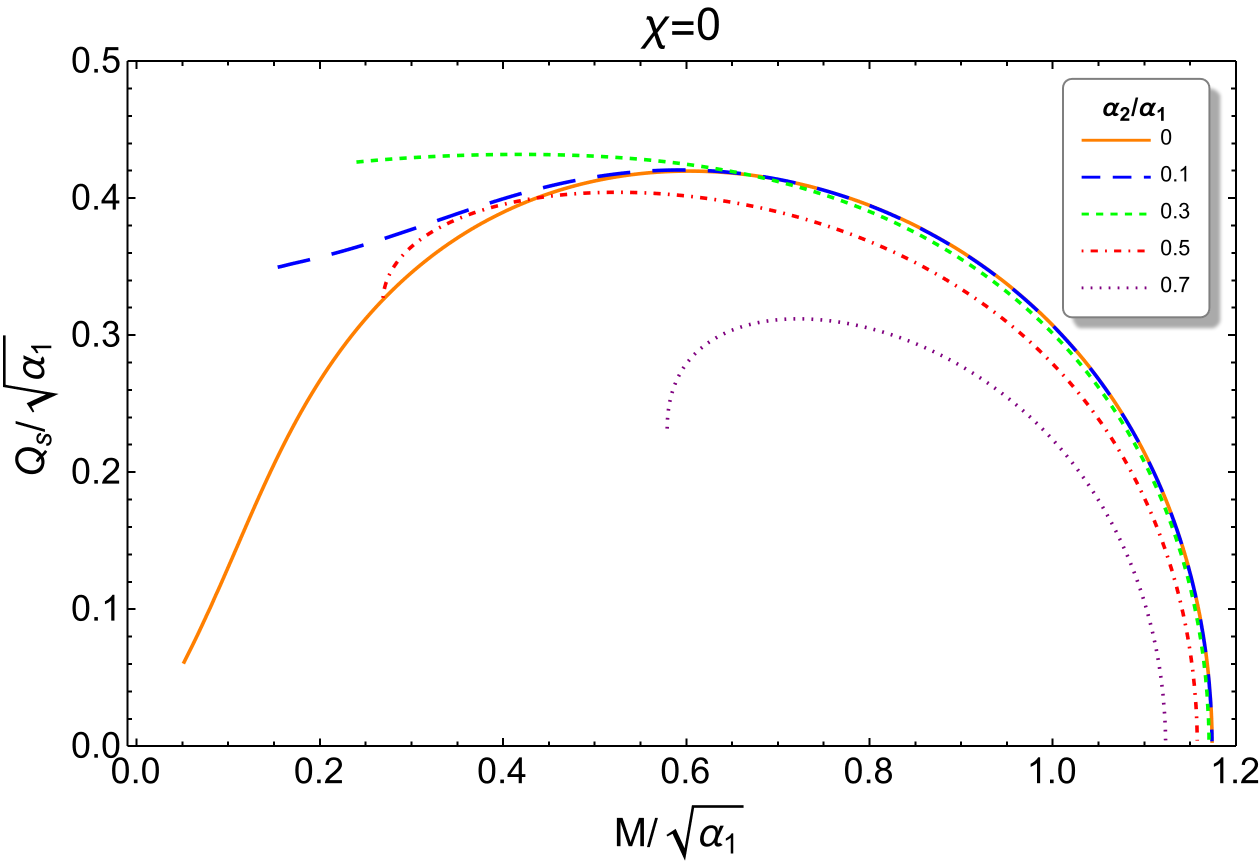}}
        % \vspace{5pt}
        % \vspace{3pt}
    \end{minipage}
    \begin{minipage}{0.4\linewidth}
        % \vspace{3pt}
        \centerline{\includegraphics[width=\textwidth]{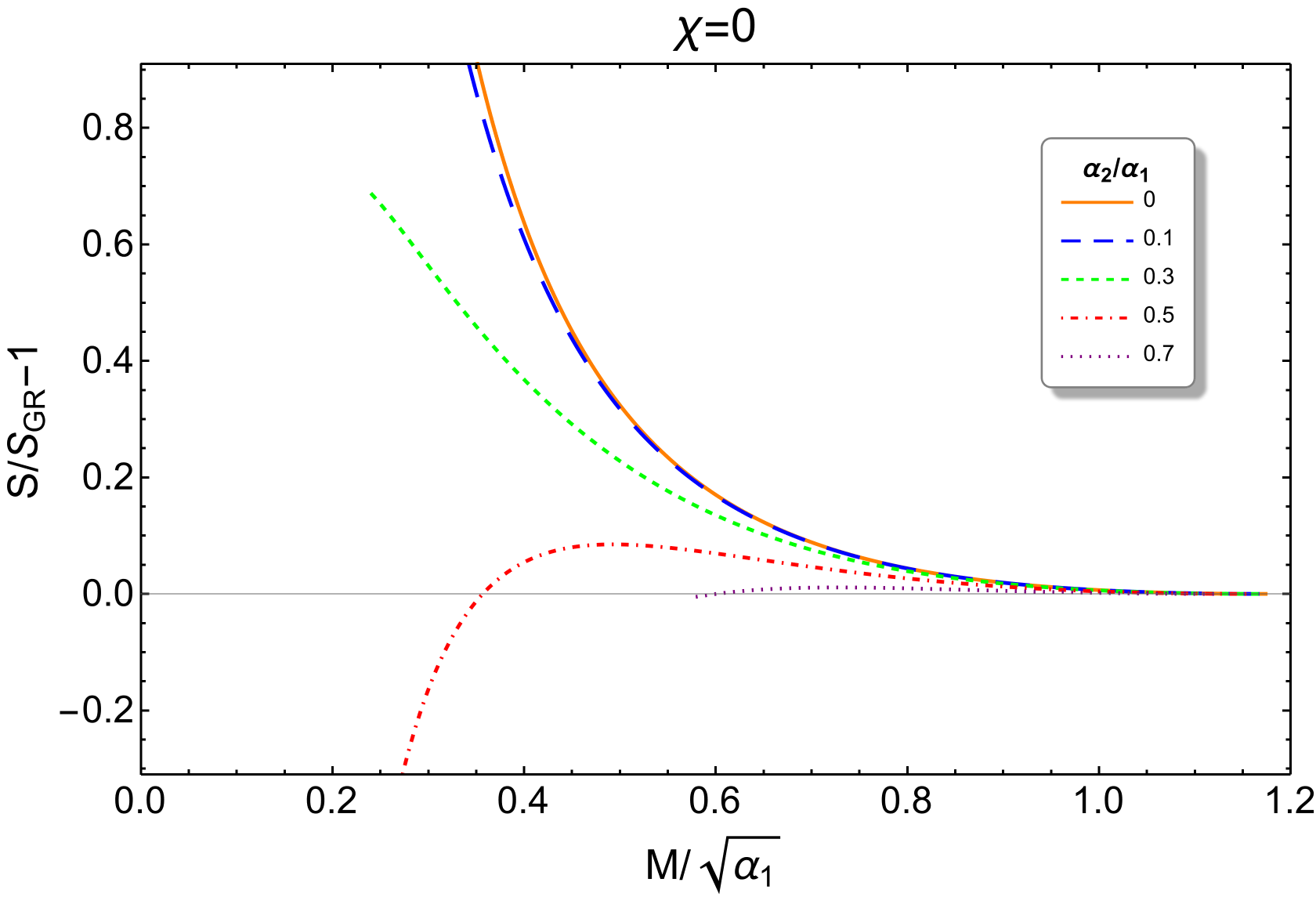}}
        % \vspace{5pt}
        % \vspace{3pt}
    \end{minipage}
    \caption{Scalar charge $Q_s/\sqrt{\alpha_1}$ (left-hand panel) and entropy $S/S_{GR} - 1$ (right-hand panel) of the scalarized solutions as a function of $M/\sqrt{\alpha_1}$.}
    \label{fig:a2=0chi=0}
\end{figure}

Fixing the spin $\chi=0$, Fig. \ref{fig:a2=0chi=0} shows both the scalar charge and the entropy of the scalarized solution.
In the no-rotating limit, the rotating scalarized solution reduces to the spherically symmetric cases, and consequently, Fig. \ref{fig:a2=0chi=0} is identical to its counterpart in \cite{eichhorn2023breaking}.
%We employ different colors and curves to differentiate the results of varying coupling ratios.
Each curve in the figure originates from an existence solution (right boundary point) and terminates at a critical solution (left boundary point).
As shown in the right panel, the entropy of a scalarized BH exceeds that of a Schwarzschild BH with the same mass, with the exception of the coupling ratios $\alpha_2 / \alpha_1 = 0.5$ and $0.7$ at small masses. 
This observation indicates that scalarized BHs are entropically favored over Schwarzschild BHs in most cases.
Furthermore, as the coupling ratio $\alpha_2 / \alpha_1$ increases, the mass window for the scalarized BHs shrinks.
Thus, we can conclude that the additional squared Gauss-Bonnet term suppresses the mass window for the scalarized BHs.
In the following sections, we provide a more detailed discussion of the effect of the quartic Gauss-Bonnet term on rotating scalarized BHs.

%%%%%%%%%%%%%%%%%%%%%%%%%%%%%%%%%%%%%%%%%%%%%%%%%%%%%%%%%%%%%%%%%%%%%%%%%%%%%%%%%%%%%%%%%%%%%%%%%%%%%%%%%%%%%%%%%%%%%%%%%%%%%%%%%%%%%%%%%%%%%%%%%%%%%%%%%%%%%%%%%%%%%%%%%%%
\subsubsection{$\alpha_2/\alpha_1=0$}

\begin{figure}[h]
    \begin{minipage}{0.45\linewidth}
        % \vspace{3pt}
        \centerline{\includegraphics[width=\textwidth]{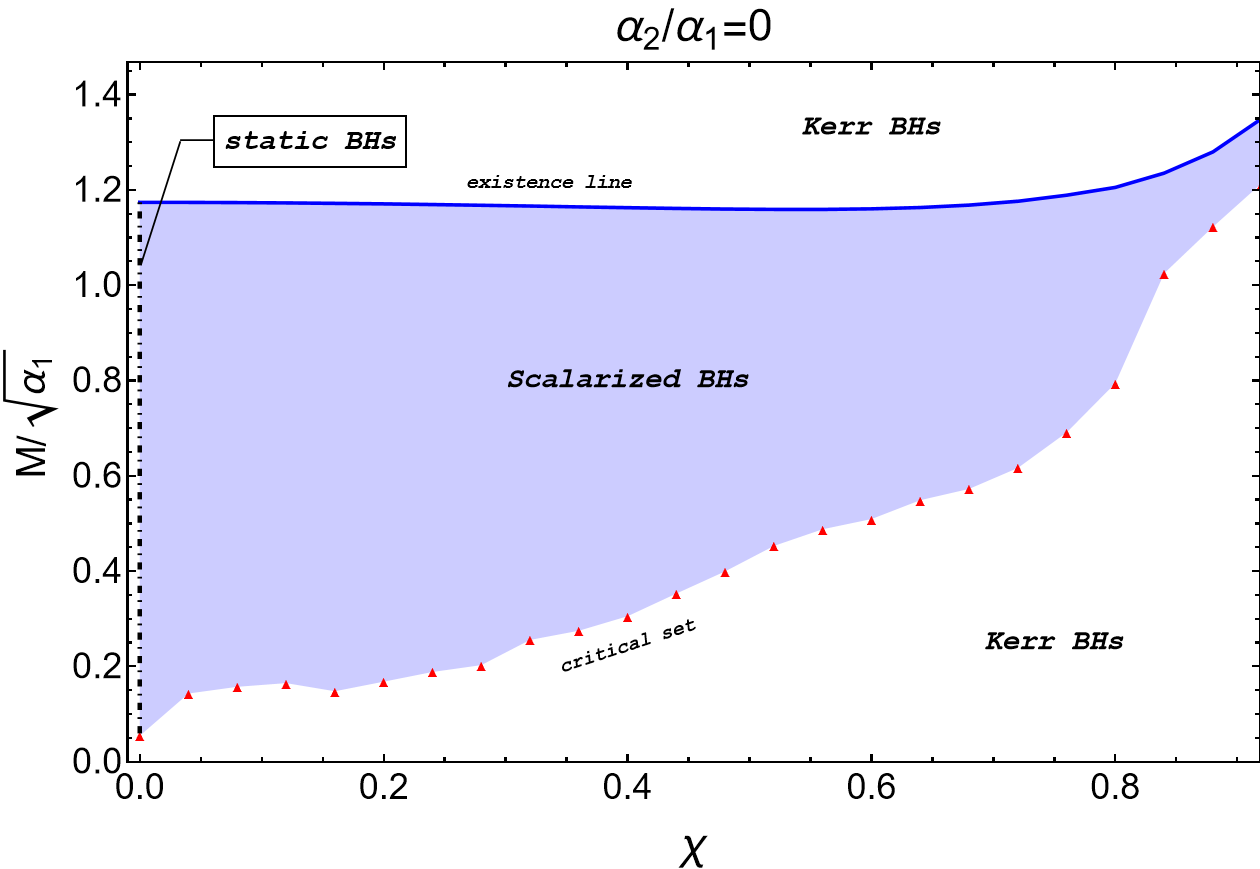}}
        % \vspace{3pt}
    \end{minipage}
   \begin{minipage}{0.45\linewidth}
       % \vspace{3pt}
         \centerline{\includegraphics[width=\textwidth]{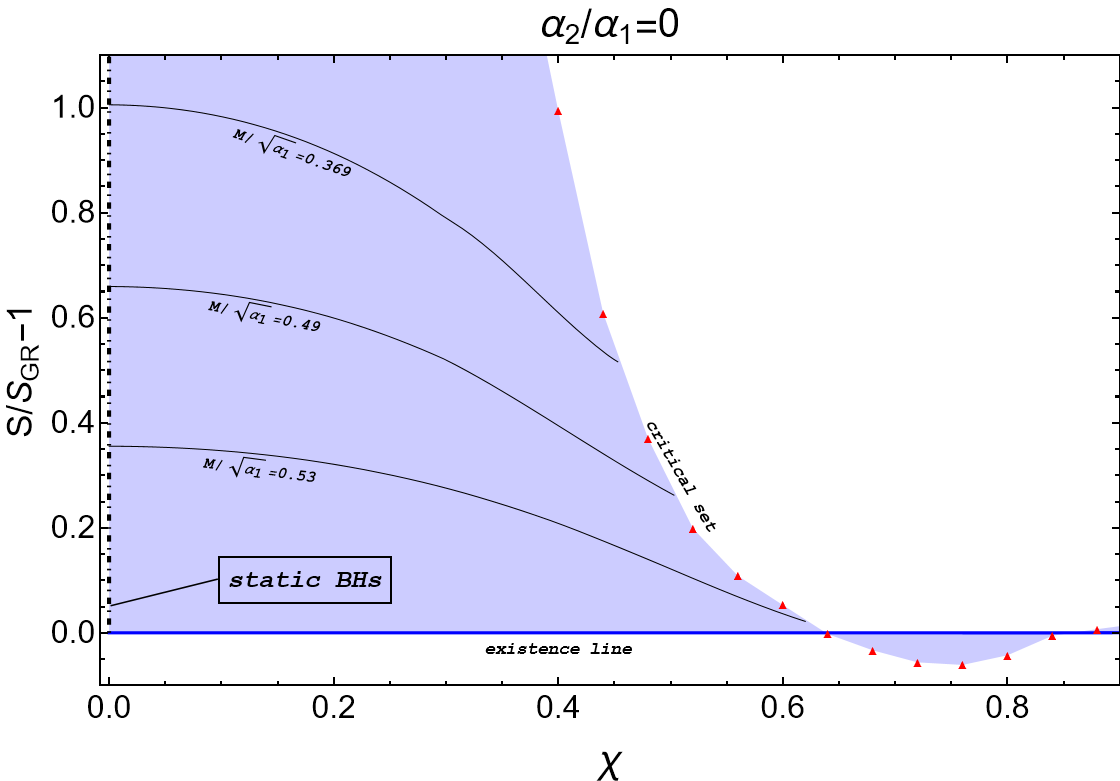}}
         % \vspace{3pt}
   \end{minipage}
    \caption{The existence domain (left-hand panel), and entropy $S/S_{GR} - 1$ (right-hand panel) of scalarized BHs for the case with the coupling parameter $\alpha_{2}/\alpha_{1}=0$.}
    \label{fig:M and S vs Chi phi2 0}
\end{figure}

Turning off the coupling parameter $\alpha_2$,  the left panel of Fig. \ref{fig:M and S vs Chi phi2 0} displays the existence domain of the scalarized BHs, while the right panel plots the entropy ratio $S/S_{GR} - 1$ as a function of the dimensionless spin $\chi$.
The existence domain (the darker shaded area) is bounded by three distinct sets of solutions: (1) the static spherically symmetric solutions labeled by dashed-dotted segment overlapping with the vertical axis where $\chi=0$; (2) the existence line (solid blue line), which corresponds to the bifurcation edge from the Kerr family; and (3) the set of critical solutions marked by red regular triangles. 
Critical solutions are a common feature in sGB models \cite{PhysRevD.90.124063,PhysRevD.54.5049,PhysRevLett.112.251102,PhysRevLett.106.151104,Kleihaus:2015aje}, and the numerical process fails to converge as the BH parameters approach the critical sets. 
This behavior can be explained by the fact that the quadratic equation of the second-order term in the near-horizon expansion of the scalar field ceases to have a real solution as the critical set is approached.
Accordingly, a uniform near-horizon expansion of the solution is no longer feasible, demonstrating that there does not exist a regular solution; for further details, see \cite{Kleihaus:2015aje,delgado2020spinning} .
Note that the static spherically symmetric solutions correspond to the solid orange curve in Fig. \ref{fig:a2=0chi=0}.
%Note that we use the same symbols to represent the three sets in the following figures.
Furthermore, the right panel compares the entropy $S$ of scalarized BHs with that of Kerr holes, $S_{GR}$, having the same mass and spin. 
The comparison shows that, for $\chi<0.64$, most scalarized solutions are thermodynamically more stable than their Kerr counterpart.

\begin{figure}[th]
    \begin{minipage}{0.45\linewidth}
        % \vspace{3pt}
        \centerline{\includegraphics[width=\textwidth]{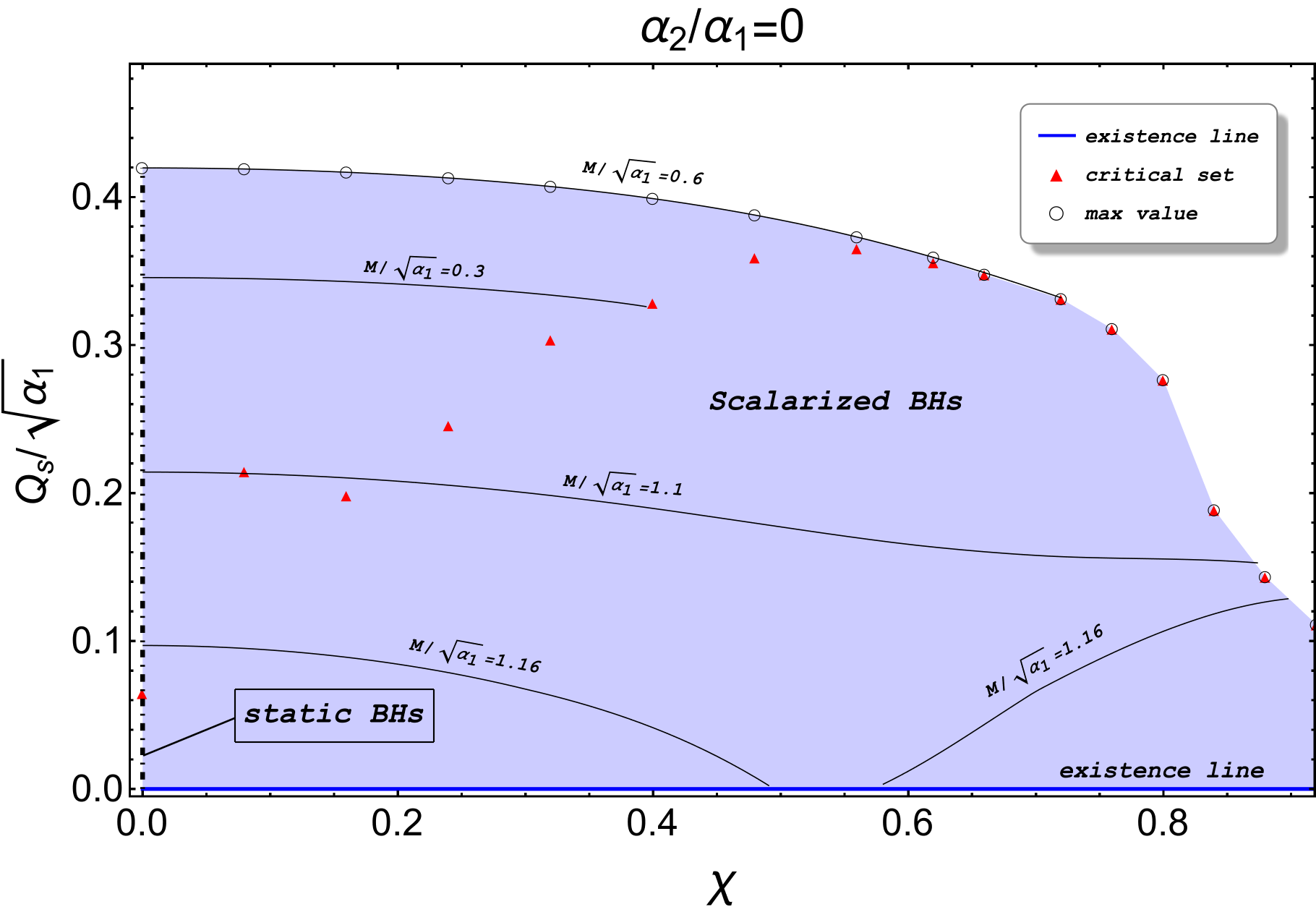}}
        % \vspace{3pt}
    \end{minipage}
    \begin{minipage}{0.45\linewidth}
        % \vspace{3pt}
        \centerline{\includegraphics[width=\textwidth]{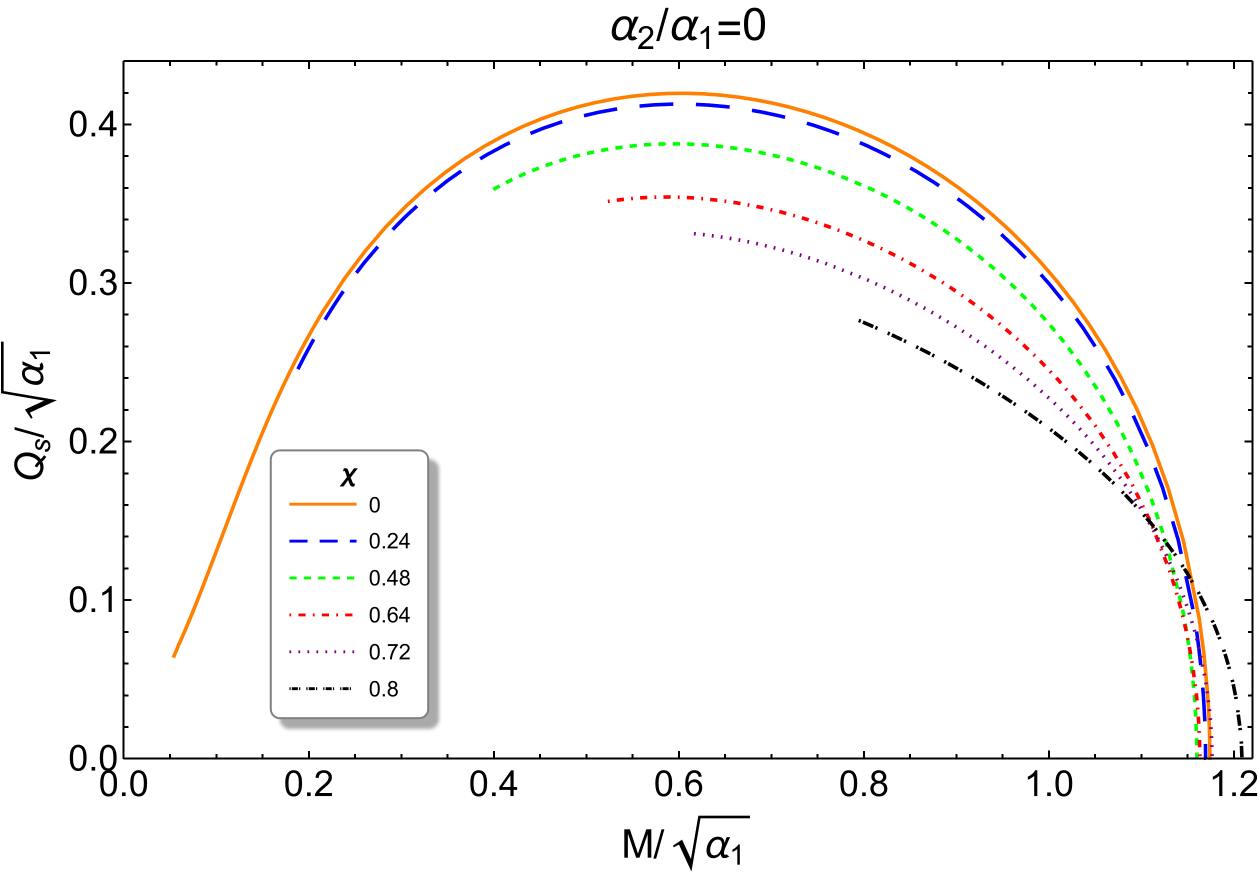}}
        % \vspace{3pt}
    \end{minipage}
    \caption{Setting $\alpha_{2}/\alpha_{1}=0$, the left panel shows scalar charge $Q_s/\sqrt{\alpha_1}$ of the scalarized solutions versus spin $\chi$, while the right panel gives scalar charge $Q_s/\sqrt{\alpha_1}$ versus ADM mass $M/\sqrt{\alpha_1}$.}
    \label{fig:Qs vs Chi phi2 0}
\end{figure}

To show the strength of the scalar field, Fig. \ref{fig:Qs vs Chi phi2 0} illustrates the scalar charge $Q_s/\sqrt{\alpha_1}$ of scalarized BHs as a function of their mass and dimensionless spin. 
In the left panel, for spins $\chi<0.72$, the scalar charge at the critical set does not coincide with the maximum values (labeled by the empty circles) that form the upper boundary. 
As the dimensionless spin $\chi \geq 0.72$, the critical sets coincide with the upper boundary.
To further examine the dependence of the scalar charge on both BH mass and spin, the right panel displays plots of $Q_{s}/\sqrt{\alpha_1}$ versus $M/\sqrt{\alpha_1}$ for various values of $\chi$.
In these curves, the left endpoint represents a critical solution (marked by a red regular triangle in the left panel), while the right endpoint corresponds to an existence solution (denoted by the solid blue line). 
As shown, for spins $\chi < 0.72$, the scalar charge initially increases with mass, reaches a maximum, and then decreases, resulting in a parabolic distribution.
As $\chi$ increases, the curves become monotonic, with the maximum scalar charge occurring at the critical solution. 
This behavior indicates that only for sufficiently large values of $\chi$($\geq 0.72$) do the scalar charge of critical solutions coincide with the maximum scalar charge.

\begin{figure}[htbp]
    \centering
    \begin{minipage}{0.45\textwidth}
        \centering
	\includegraphics[width=\textwidth]{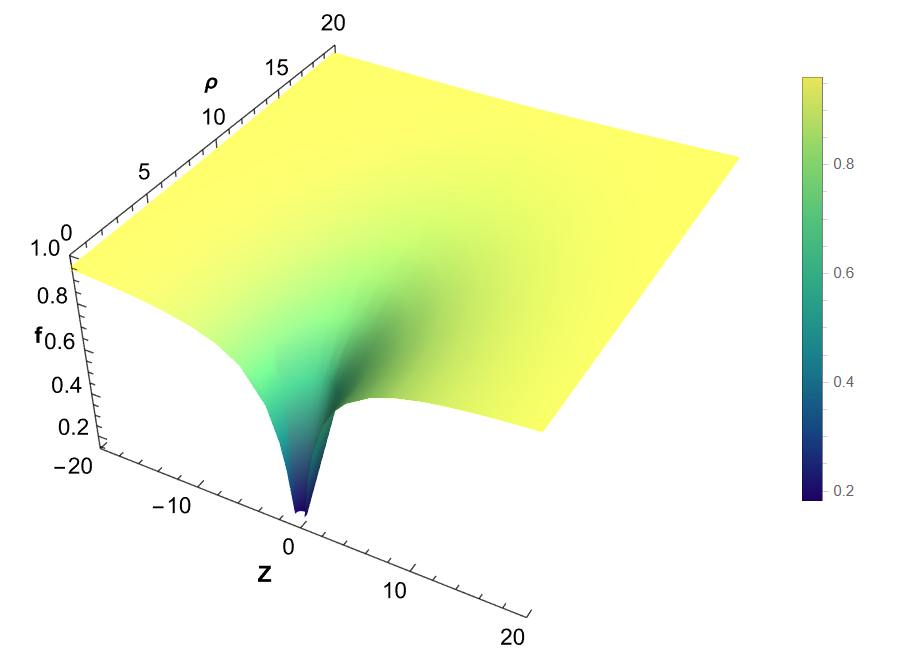}
    \end{minipage}
    \hspace{0.5cm}
    \begin{minipage}{0.45\textwidth}
	\centering
	\includegraphics[width=\textwidth]{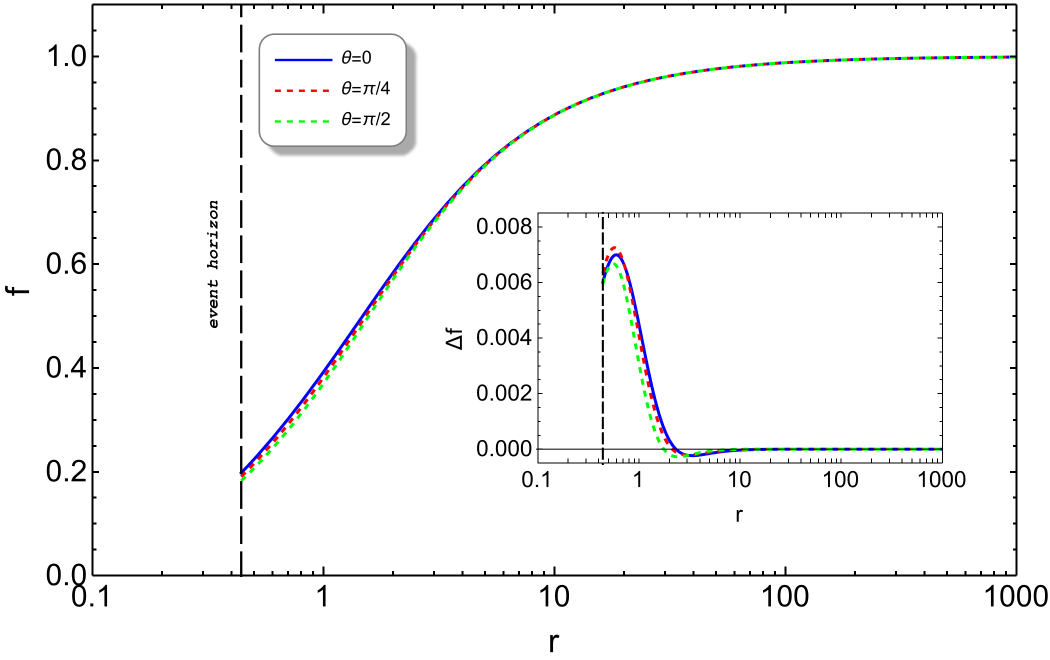}
    \end{minipage}
    
    \begin{minipage}{0.45\textwidth}
        \centering
	\includegraphics[width=\textwidth]{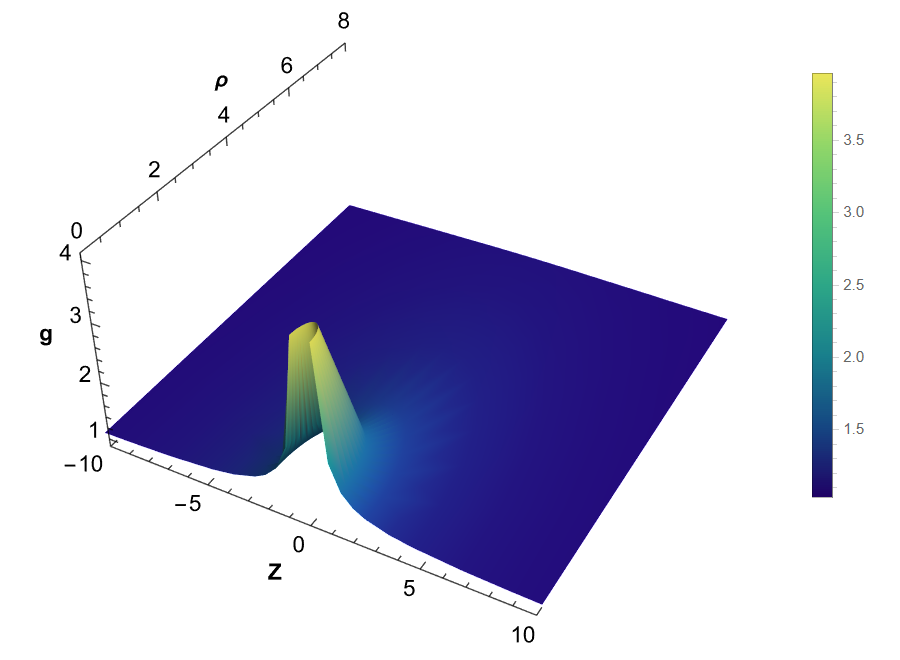}
    \end{minipage}
    \hspace{0.5cm}
    \begin{minipage}{0.45\textwidth}
	\centering
	\includegraphics[width=\textwidth]{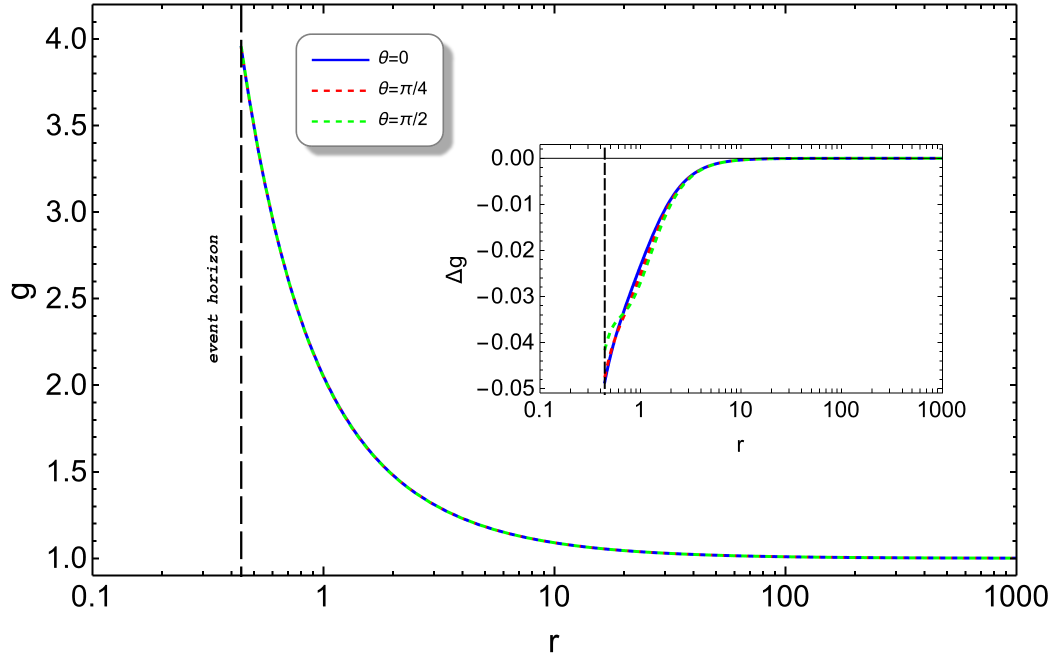}
    \end{minipage}
    
    \caption{Metric functions $f$ and $g$ for scalarized rotating BH solution with the parameters $\chi$ = 0.5, $r_H$ = 0.44 (dashed black line) and $\alpha_2$ = 0. The deviations between the scalarized BH and the Kerr BH are described by $\Delta f = f - f_{Kerr}$ and $\Delta g = g - g_{Kerr}$.}
    \label{fig:3D 2D metric functions}
\end{figure}

\begin{figure}[htbp]
    \centering
    \begin{minipage}{0.45\textwidth}
        \centering
	\includegraphics[width=\textwidth]{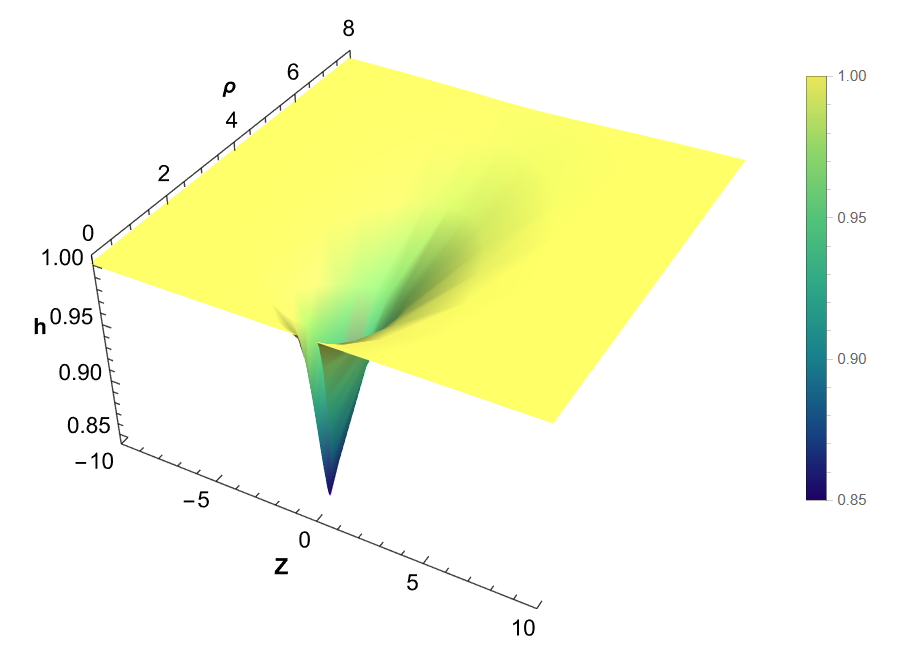}
    \end{minipage}
    \hspace{0.5cm}
    \begin{minipage}{0.45\textwidth}
	\centering
	\includegraphics[width=\textwidth]{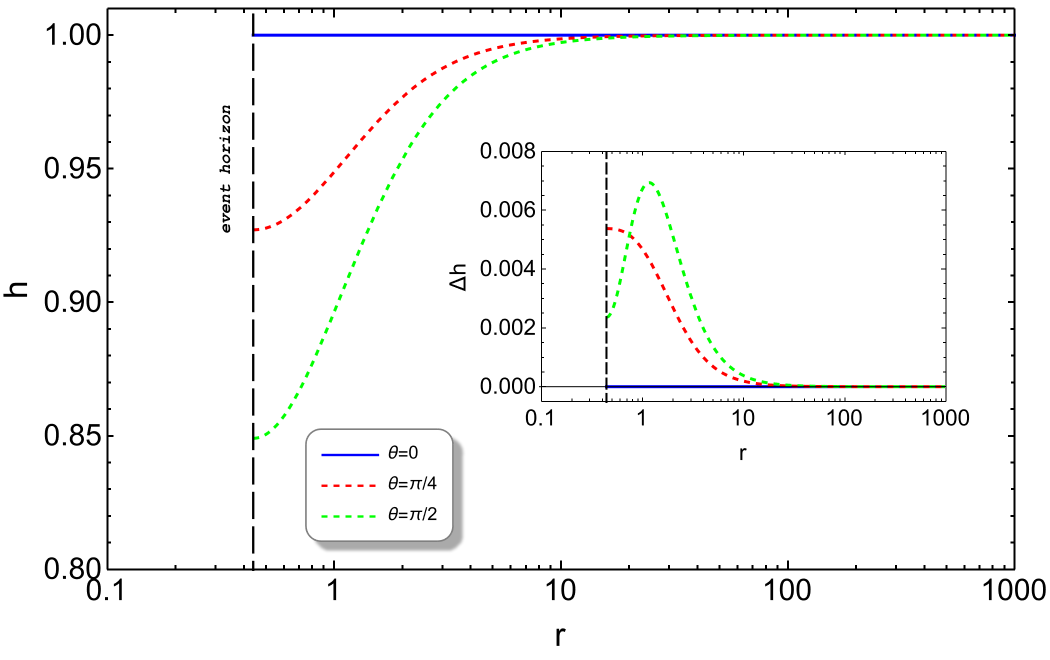}
    \end{minipage}
    
    \begin{minipage}{0.45\textwidth}
        \centering
	\includegraphics[width=\textwidth]{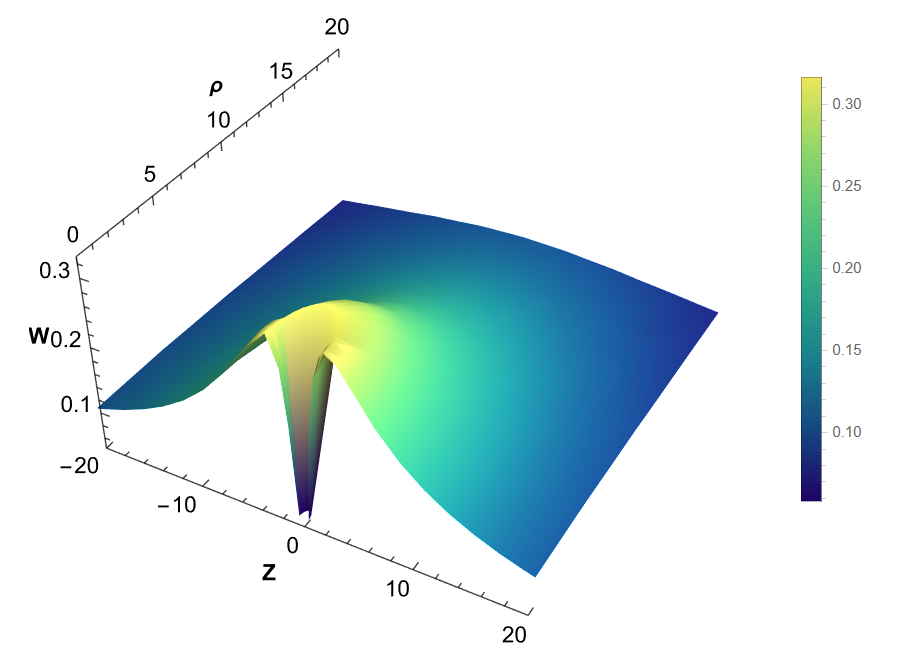}
    \end{minipage}
    \hspace{0.5cm}
    \begin{minipage}{0.45\textwidth}
	\centering
	\includegraphics[width=\textwidth]{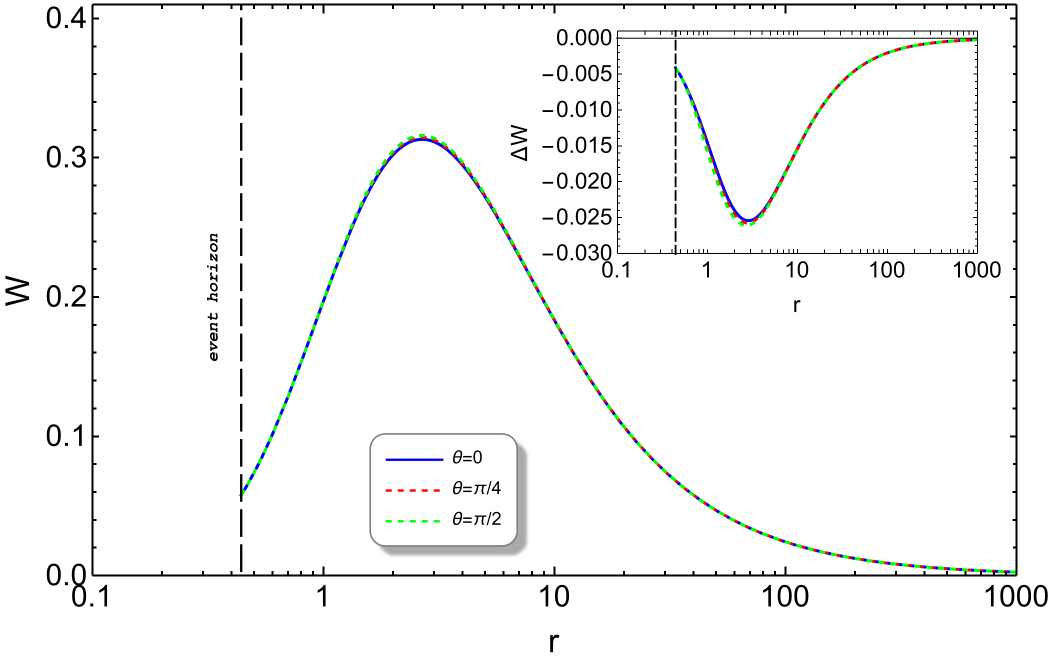}
    \end{minipage}

    \begin{minipage}{0.45\textwidth}
        \centering
	\includegraphics[width=\textwidth]{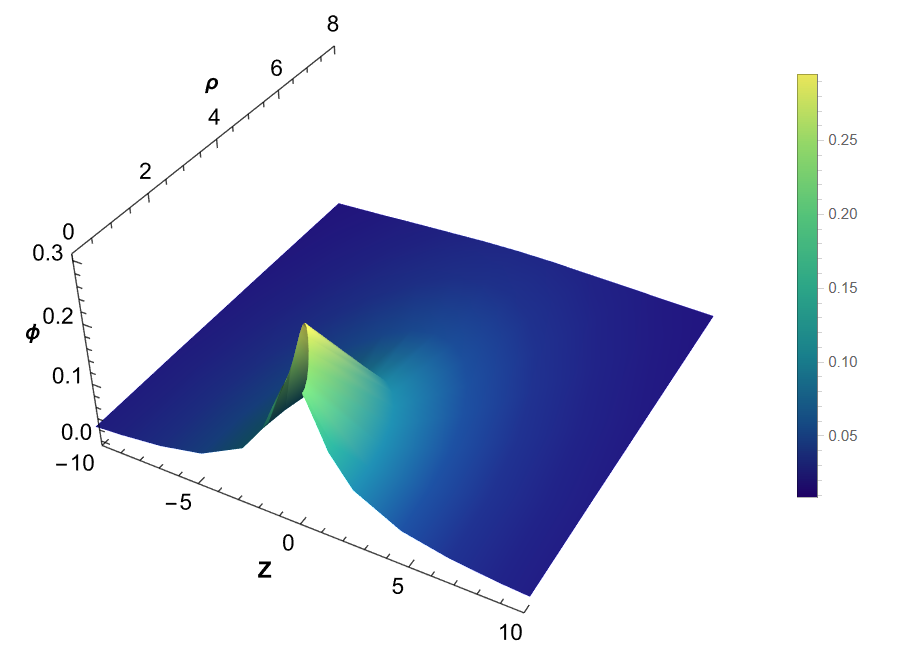}
    \end{minipage}
    \hspace{0.5cm}
    \begin{minipage}{0.45\textwidth}
	\centering
	\includegraphics[width=\textwidth]{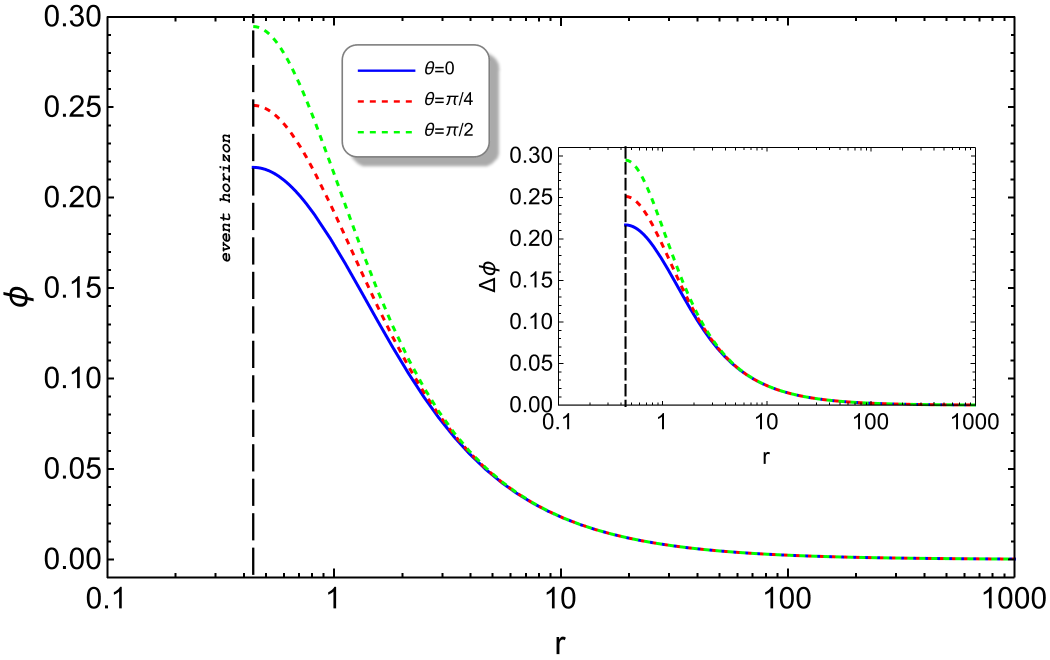}
    \end{minipage}

    \begin{minipage}{0.45\textwidth}
        \centering
	\includegraphics[width=\textwidth]{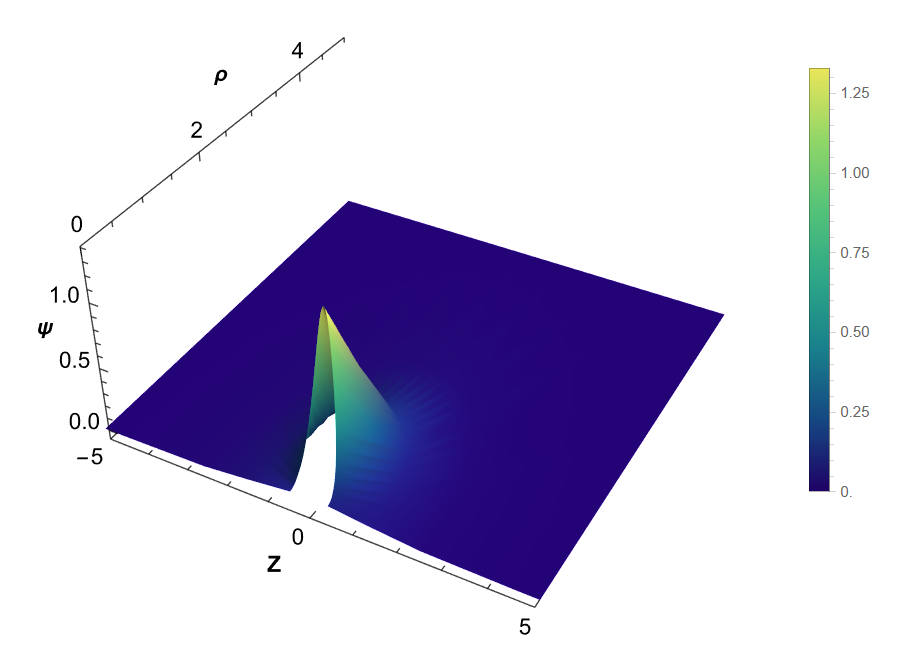}
    \end{minipage}
    \hspace{0.5cm}
    \begin{minipage}{0.45\textwidth}
	\centering
	\includegraphics[width=\textwidth]{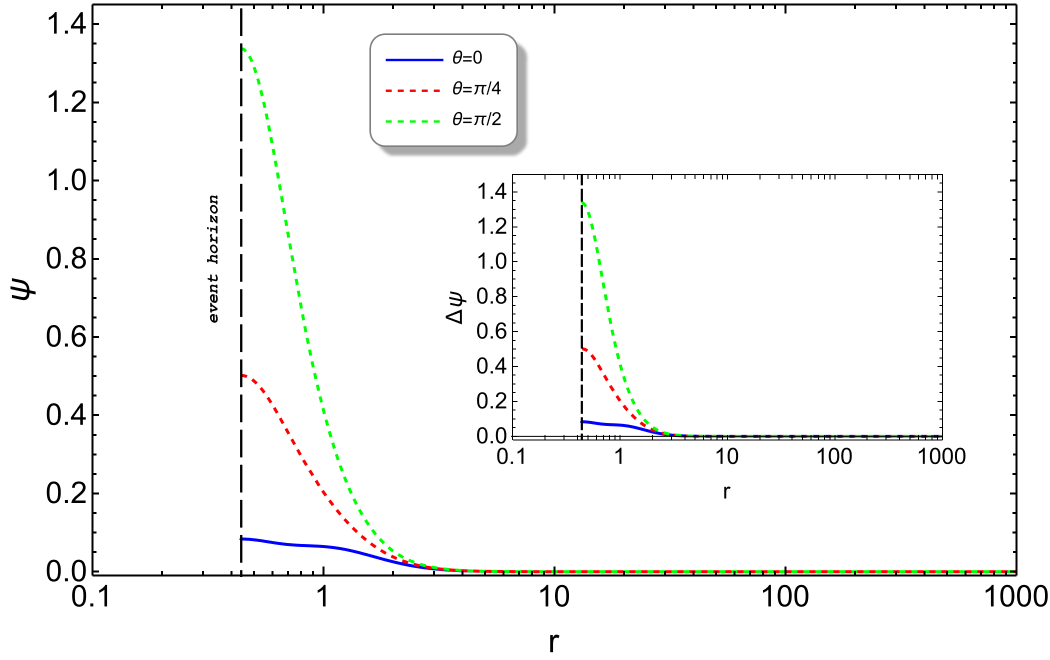}
    \end{minipage}
    \caption{Metric functions $h$, $W$ and scalar fields for scalarized rotating BH solution with the same parameters as in Fig. \ref{fig:3D 2D metric functions}. The deviations between the scalarized BH and the Kerr BH are described by $\Delta h = h - h_{Kerr}$ and $\Delta W = W - W_{Kerr}$ .}
    \label{fig:3D 2D scalar fields}
\end{figure}

Using the parameters $\chi=0.5$, $r_H = 0.44$, we obtained a numerical solution with the BH mass $M= 1.04724$ and the scalar charge $Q_s=0.234718$.
Figs. \ref{fig:3D 2D metric functions} and \ref{fig:3D 2D scalar fields} present the corresponding metric function and scalar fields.
In these figures, the left columns display three-dimensional (3D) plots, while the right columns show two-dimensional (2D) plots of the functions as a function of the radial variable for three different angular values.
For the 3D plots, the axes are defined by $\rho = r \sin{\theta}$ and $Z = r \cos{\theta}$ (with $r \geq r_H$).
Additionally, we depict the deviations between the scalarized rotating BH and the Kerr BH with the same $\chi$ and $r_H$ in these figures.
From Figs. \ref{fig:3D 2D metric functions} and \ref{fig:3D 2D scalar fields}, one can see that our numerical solutions exhibit smooth profiles, which leads to finite curvature invariants in the full domain of integration, and our scalarized rotating BH solution is asymptotically flat and has scalar hair.

%%%%%%%%%%%%%%%%%%%%%%%%%%%%%%%%%%%%%%%%%%%%%%%%%%%%%%%%%%%%%%%%%%%%%%%%%%%%%%%%%%%%%%%%%%%%%%%%%%%%%%%%%%%%%%%%%%%%%%%%%%%%%%%%%%%%%%%%%

\subsubsection{$\alpha_2/\alpha_1=0.3$}
\begin{figure}[h!]
    \begin{minipage}{0.45\linewidth}
        \centerline{\includegraphics[width=\textwidth]{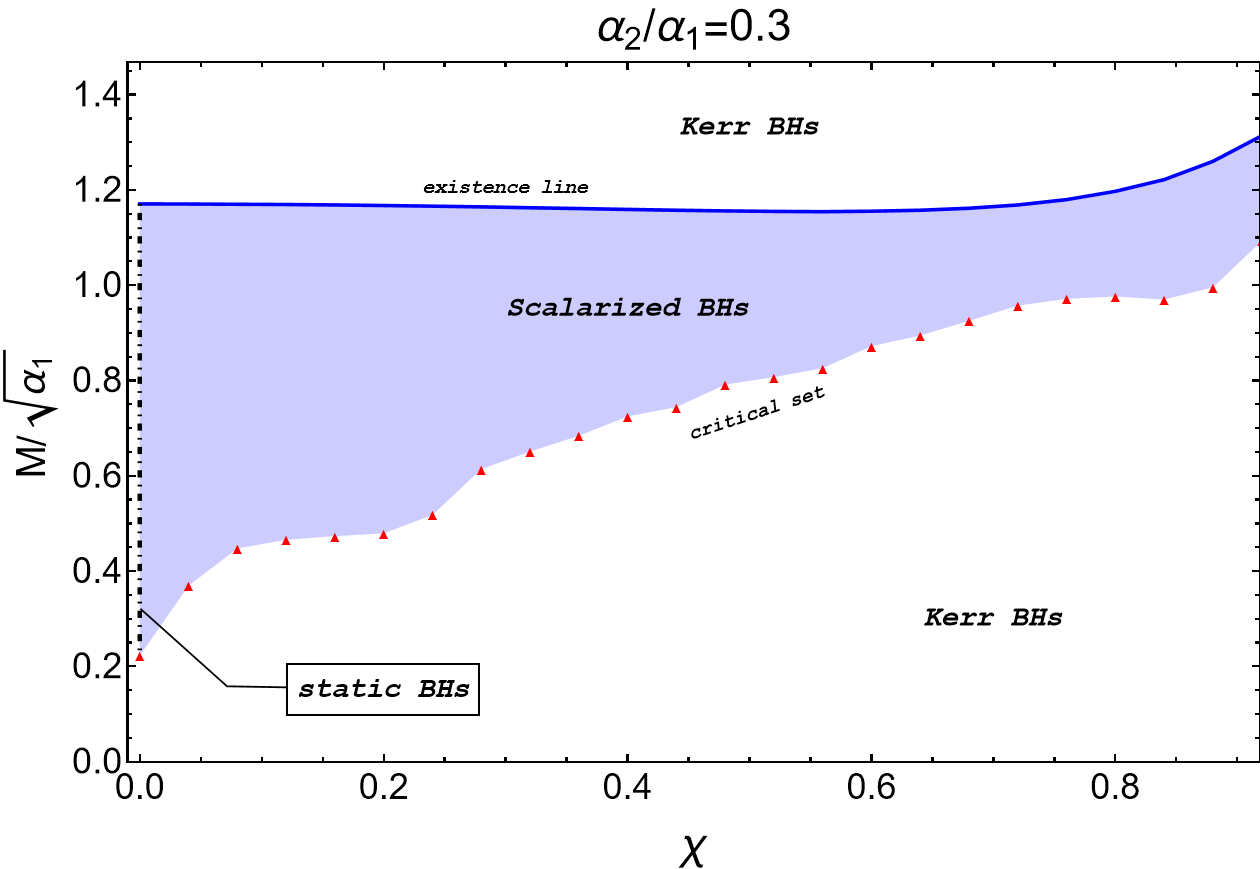}}
    \end{minipage}
    \begin{minipage}{0.45\linewidth}
        \centerline{\includegraphics[width=\textwidth]{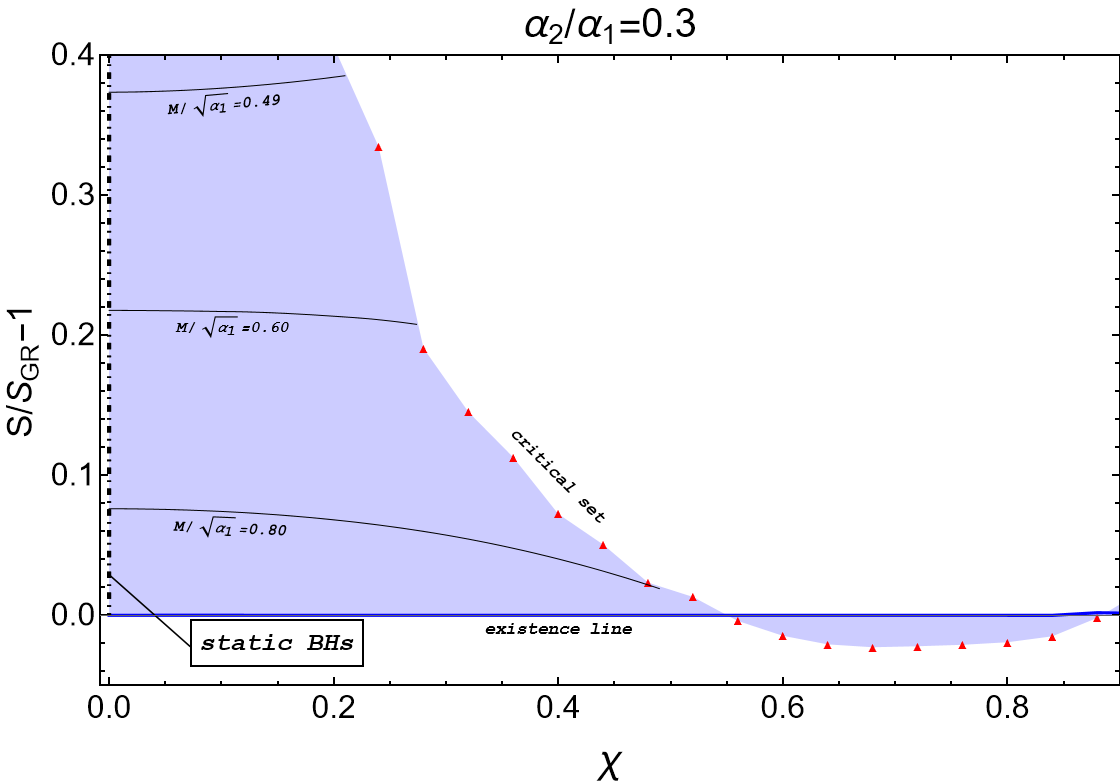}}
    \end{minipage}
    \caption{ADM mass $M/\sqrt{\alpha_1}$ (left-hand panel) and entropy $S/S_{GR} - 1$ (right-hand panel) as functions of dimensionless spin $\chi$ with the coupling parameters $\alpha_{2}/\alpha_{1}=0.3$.}
    \label{fig:M and S vs Chi phi2 0.3}
\end{figure}

To disclose the impact of additional term $\mathcal{G}^2$ on scalarization, we set the coupling parameter to be $\alpha_{2}/\alpha_{1}=0.3$ without loss of generality.
In the left panel of Fig. \ref{fig:M and S vs Chi phi2 0.3}, we give the existence domain of scalarized BHs, parametrized by mass $M/ \sqrt[]{\alpha_1}$ and dimensionless spin $\chi$.
The domain of scalarized BHs (darker shaded area) is bounded by three sets of solutions: the static BHs (dash-dotted line), the existence line (solid blue line), and the critical sets (red regular triangles). 
The static solutions here correspond to the green dished curve in Fig. \ref{fig:a2=0chi=0}.
Comparing Fig. \ref{fig:M and S vs Chi phi2 0.3} with Fig. \ref{fig:M and S vs Chi phi2 0}, we see that the region of scalarized BHs becomes compressed as $\alpha_2$ increases, and the scalarization is suppressed for lower massive BHs.
We also calculated the entropy $S$ of the scalarized BHs, and compared it with that of Kerr BHs $S_{GR}$ with the same mass and spin.
The right panel of Fig. \ref{fig:M and S vs Chi phi2 0.3} displays the relative entropy $S/S_{GR} - 1$.
In the region where $\chi\lesssim 0.55$, we find that $S/S_{GR}-1>0$, indicating that scalarized solutions are more favored by thermodynamics. 

\begin{figure}[htbp]
    \begin{minipage}{0.45\linewidth}
    % \centerline{\includegraphics[width=0.45\linewidth]{Qs_vs_Chi_Phi2_0.3.png}}
         \centerline{\includegraphics[width=\textwidth]{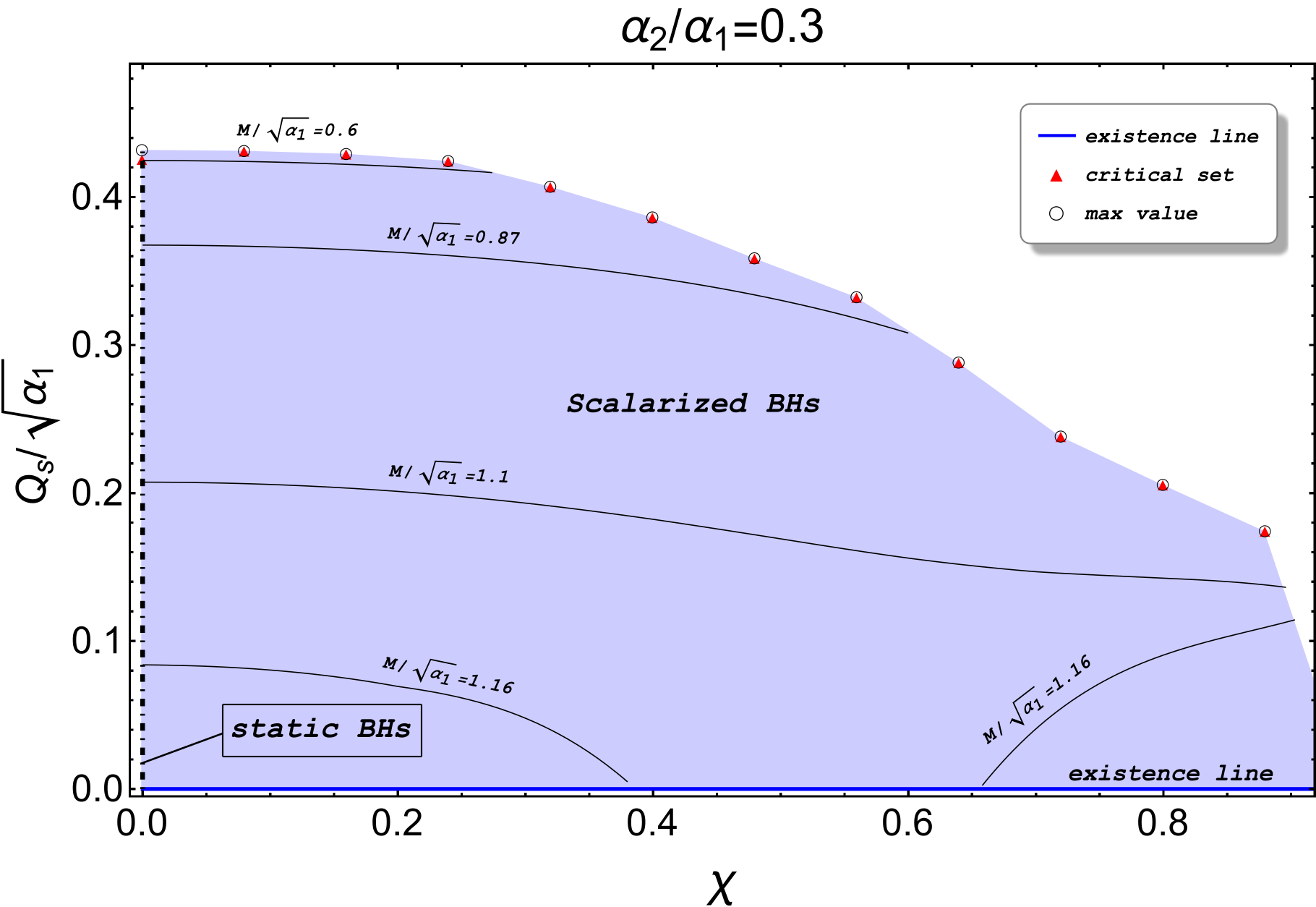}}
    \end{minipage}
    \begin{minipage}{0.45\linewidth}
        \centerline{\includegraphics[width=\textwidth]{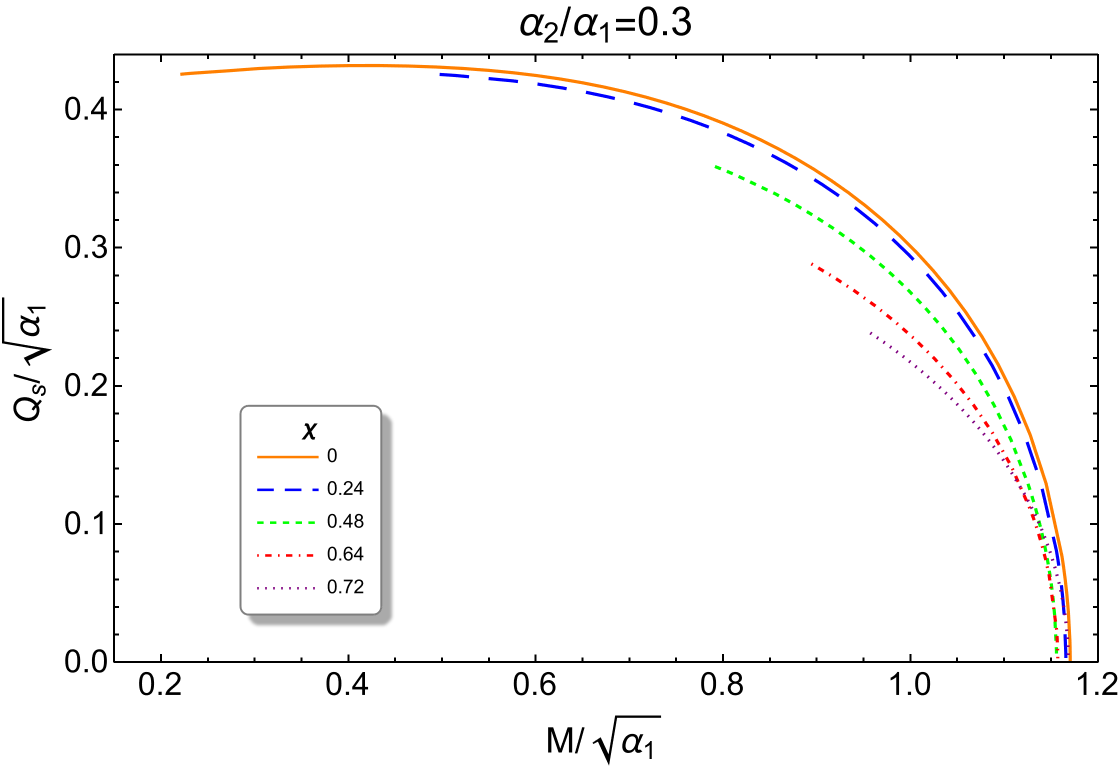}}
    \end{minipage}
    \caption{Scalar charge $Q_s/\sqrt{\alpha_1}$ of the scalarized solutions versus $\chi$ (left panel), and versus $M/\sqrt{\alpha_1}$ (right panel).}
    \label{fig:Qs vs Chi phi2 0.3}
\end{figure}

Fig. \ref{fig:Qs vs Chi phi2 0.3} illustrates how the scalar charge varies with BH mass and spin.
In the left panel, the dashed-dotted line corresponds to the orange solid line in the right panel, and it also matches the green dashed curve in the left panel of Fig. \ref{fig:a2=0chi=0}.
For BHs with $\chi<0.08$, the scalar charge of critical solutions (marked by empty circles) serves as the upper bound; however, in the low-spin regime $\chi < 0.08$, the maximum scalar charge shifts from critical solutions to intermediate parametric regions.
In the right panel, all the curves originate from critical solutions and terminate at existence solutions (characterized by a vanishing scalar charge $Q_s/\sqrt{\alpha_1}=0$).
The curve with $\chi<0.08$ follow parabolic behavior, whereas those for $\chi\geq 0.08$ follow monotonic shapes.
Comparing Fig. \ref{fig:Qs vs Chi phi2 0.3} with Fig. \ref{fig:Qs vs Chi phi2 0}, we find that the region of scalarized BHs reduces when the coupling ratio $\alpha_2/\alpha_1$ increases. 
This finding indicates that the additional term $\mathcal{G}^2$ exerts an inhibitory effect on scalarization.
%In particular, both cases share two common characteristics: (1) as $\chi$ increases, the scalar charge $Q_s/\sqrt{\alpha_1}$ tends to decrease, suggesting that static scalarized BHs exhibit a wider range of scalar charge values compared to most spinning scalarized BHs; (2) in most cases, scalarized BHs are entropically favored over Kerr BHs.

\begin{figure}[htbp]
    \centering
    \begin{minipage}{0.4\textwidth}
        \centering
	\includegraphics[width=\textwidth]{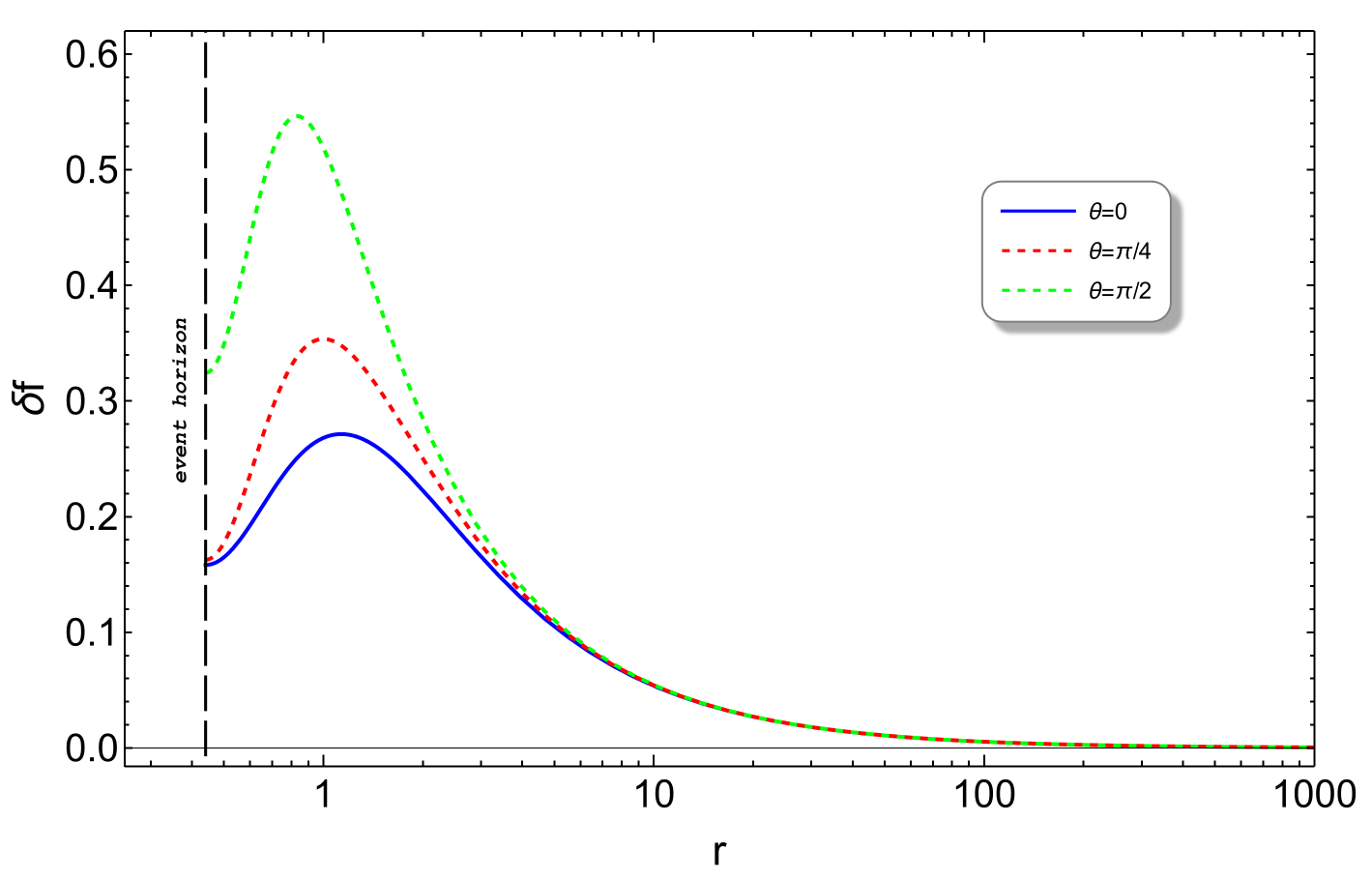}
    \end{minipage}
    \hspace{0.5cm}
    \begin{minipage}{0.4\textwidth}
	\centering
	\includegraphics[width=\textwidth]{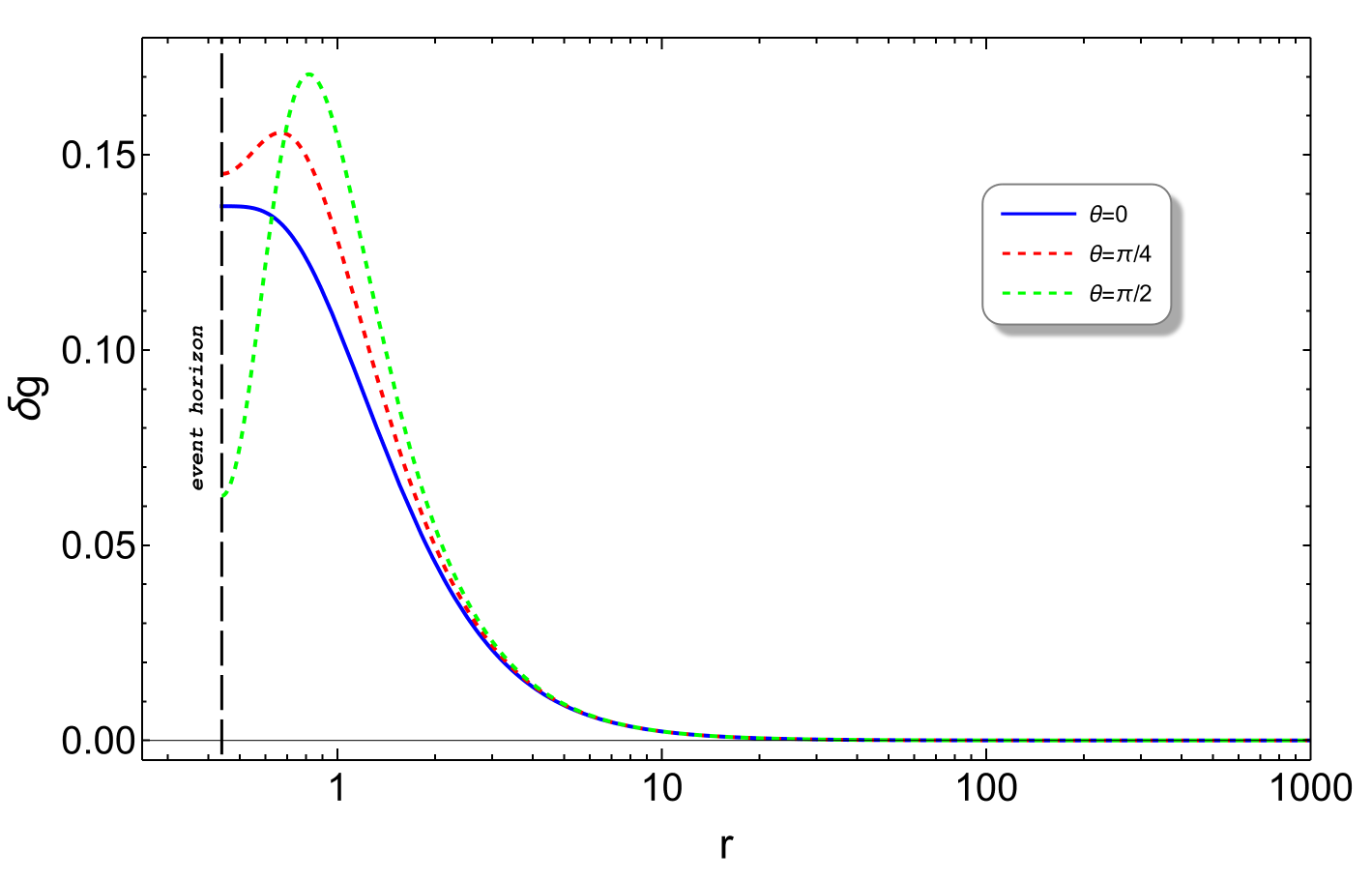}
    \end{minipage}
    
    \begin{minipage}{0.4\textwidth}
        \centering
	\includegraphics[width=\textwidth]{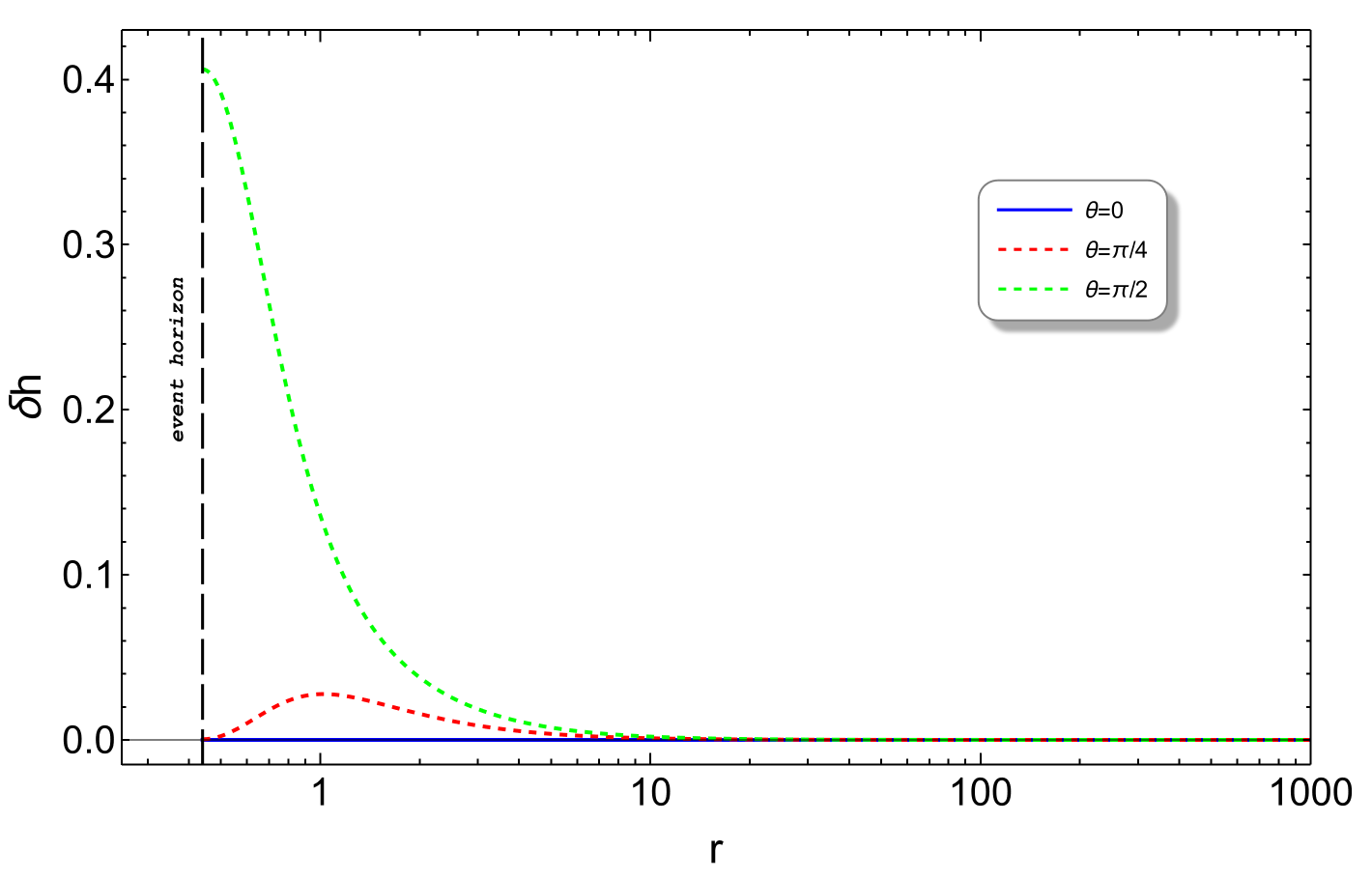}
    \end{minipage}
    \hspace{0.5cm}
    \begin{minipage}{0.4\textwidth}
	\centering
	\includegraphics[width=\textwidth]{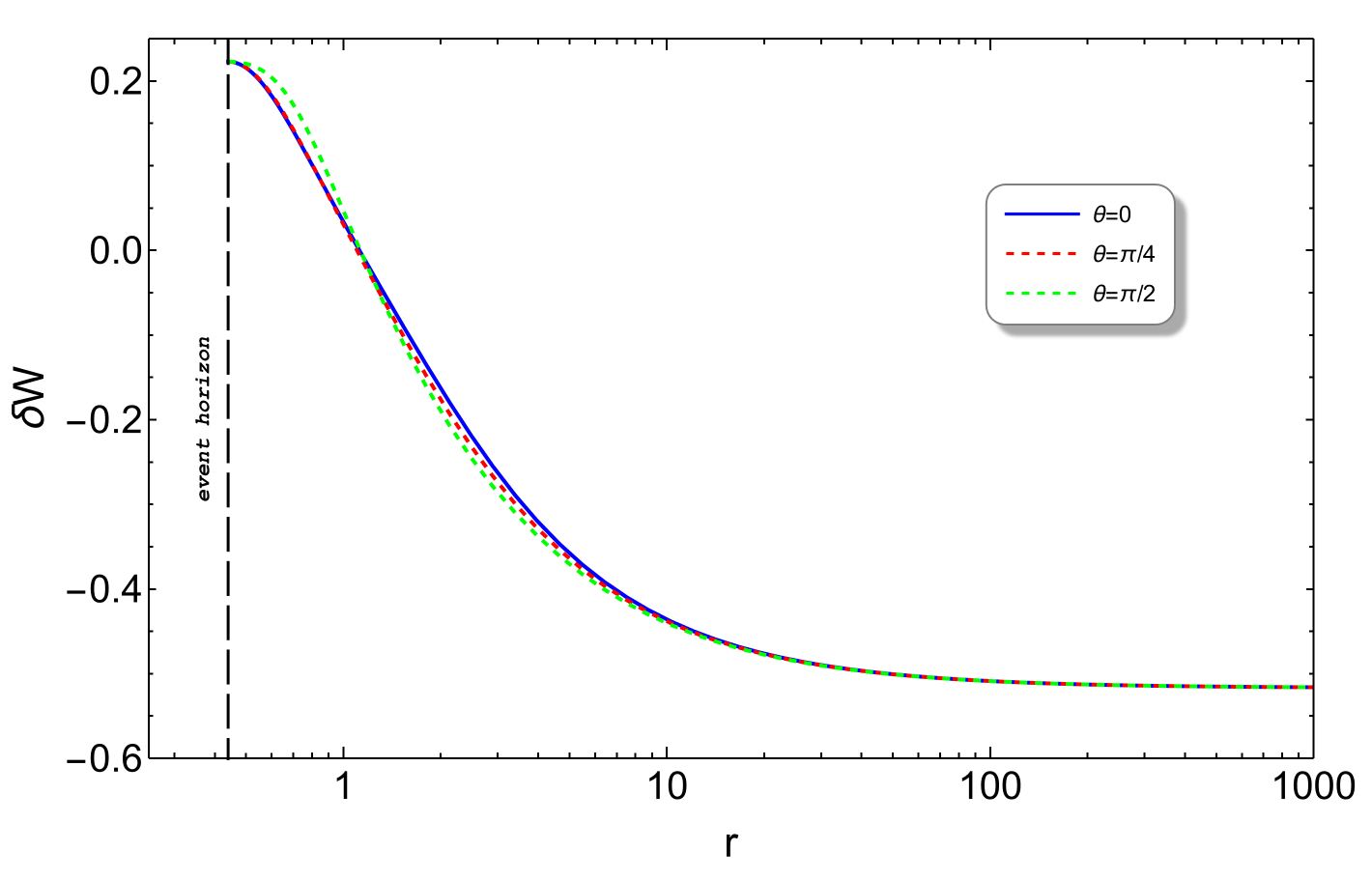}
    \end{minipage}

    \begin{minipage}{0.4\textwidth}
        \centering
	\includegraphics[width=\textwidth]{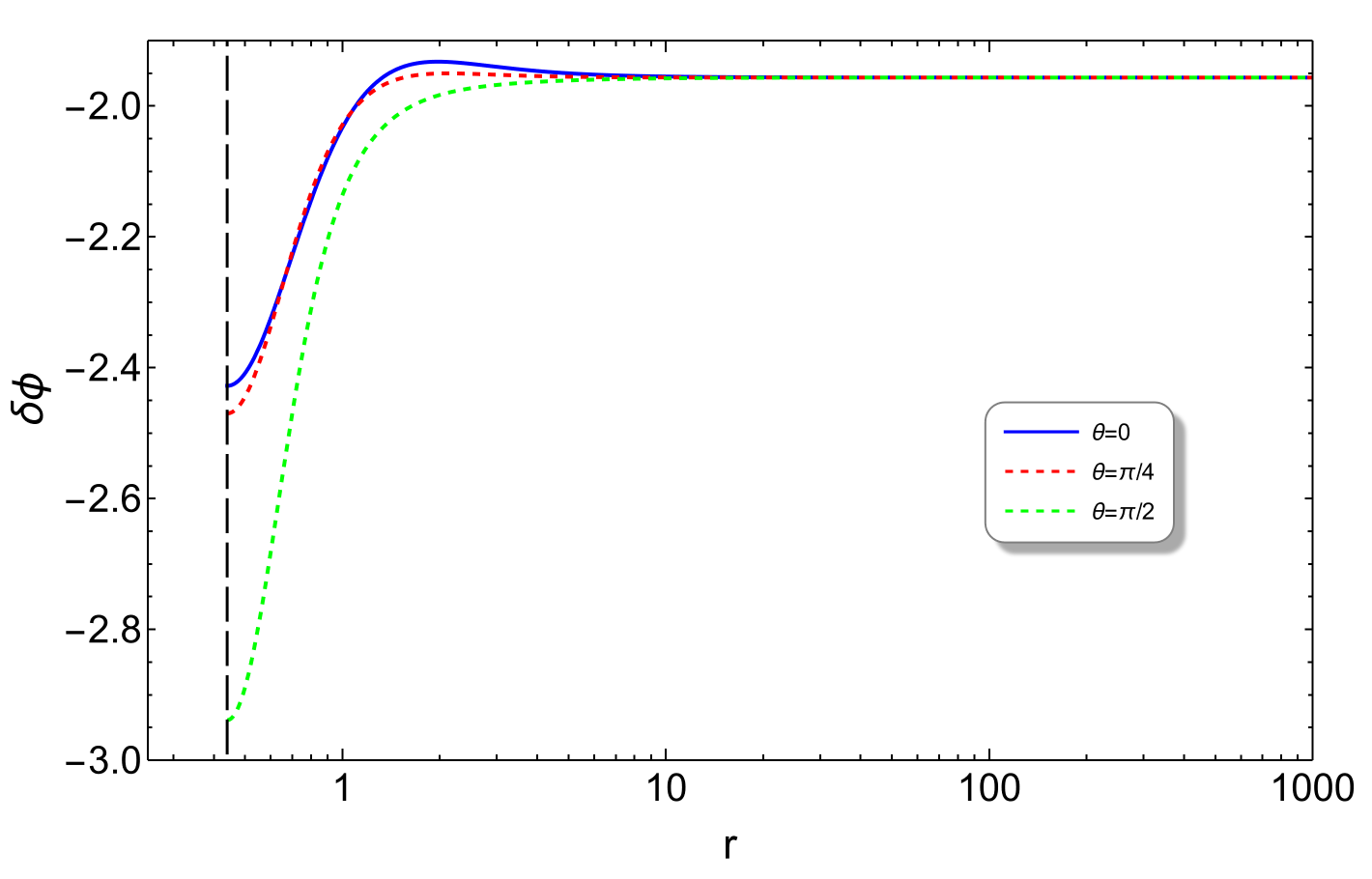}
    \end{minipage}
    \hspace{0.5cm}
  \begin{minipage}{0.4\textwidth}
	\centering
	\includegraphics[width=\textwidth]{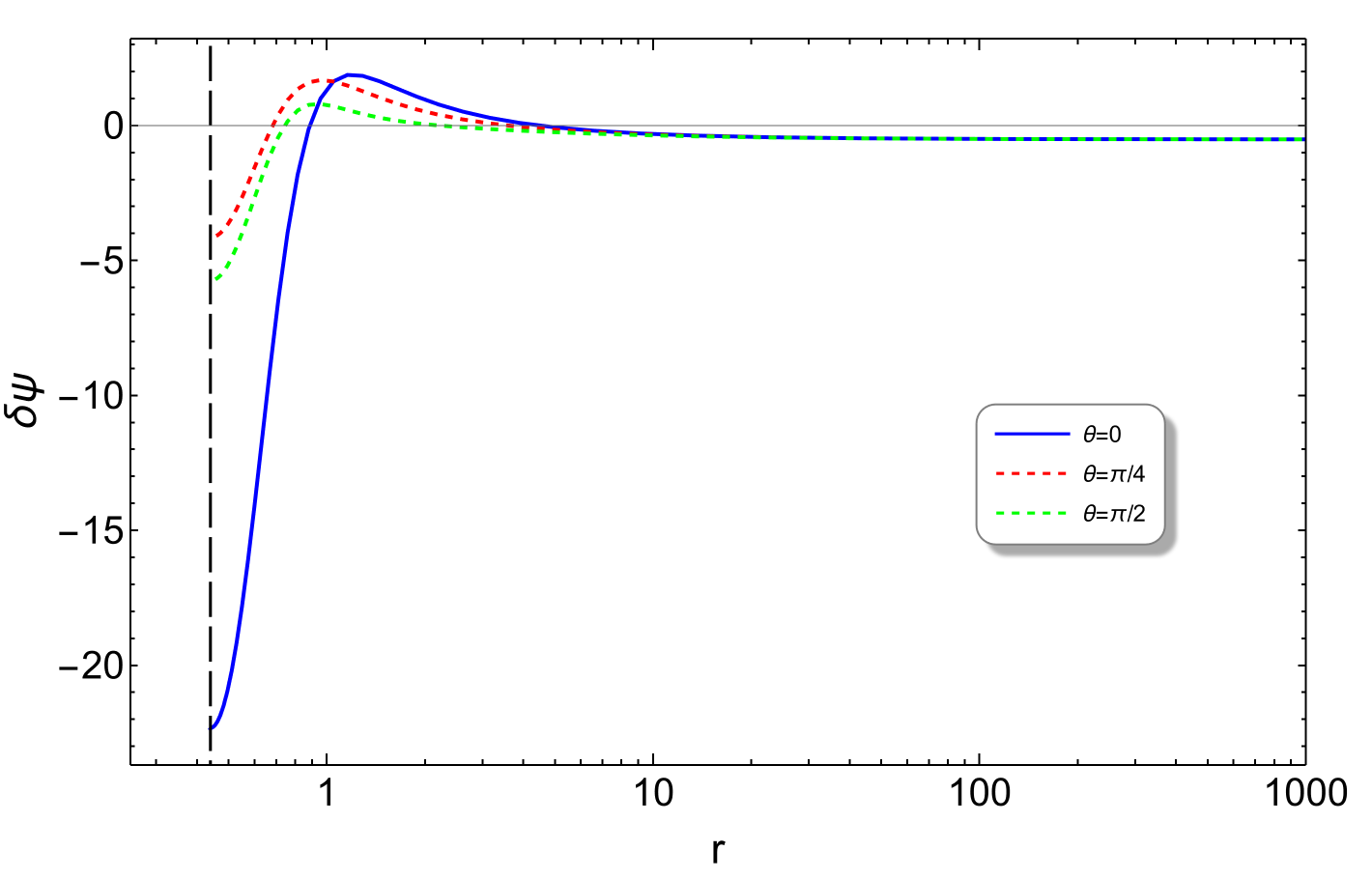}
    \end{minipage}

    \caption{Comparison of the metric functions $f,g,h,W$ and the scalar fields $\phi$ and $\psi$ for the scalarized rotating BH solutions with coupling ratio $\alpha_2/\alpha_1=0$ and $\alpha_2/\alpha_1=0.3$, using the same parameters as Fig. \ref{fig:3D 2D metric functions}.}
    \label{fig:error}
\end{figure}

\begin{figure}[htbp]
    \centering
    \begin{minipage}{0.4\textwidth}
        \centering
	\includegraphics[width=\textwidth]{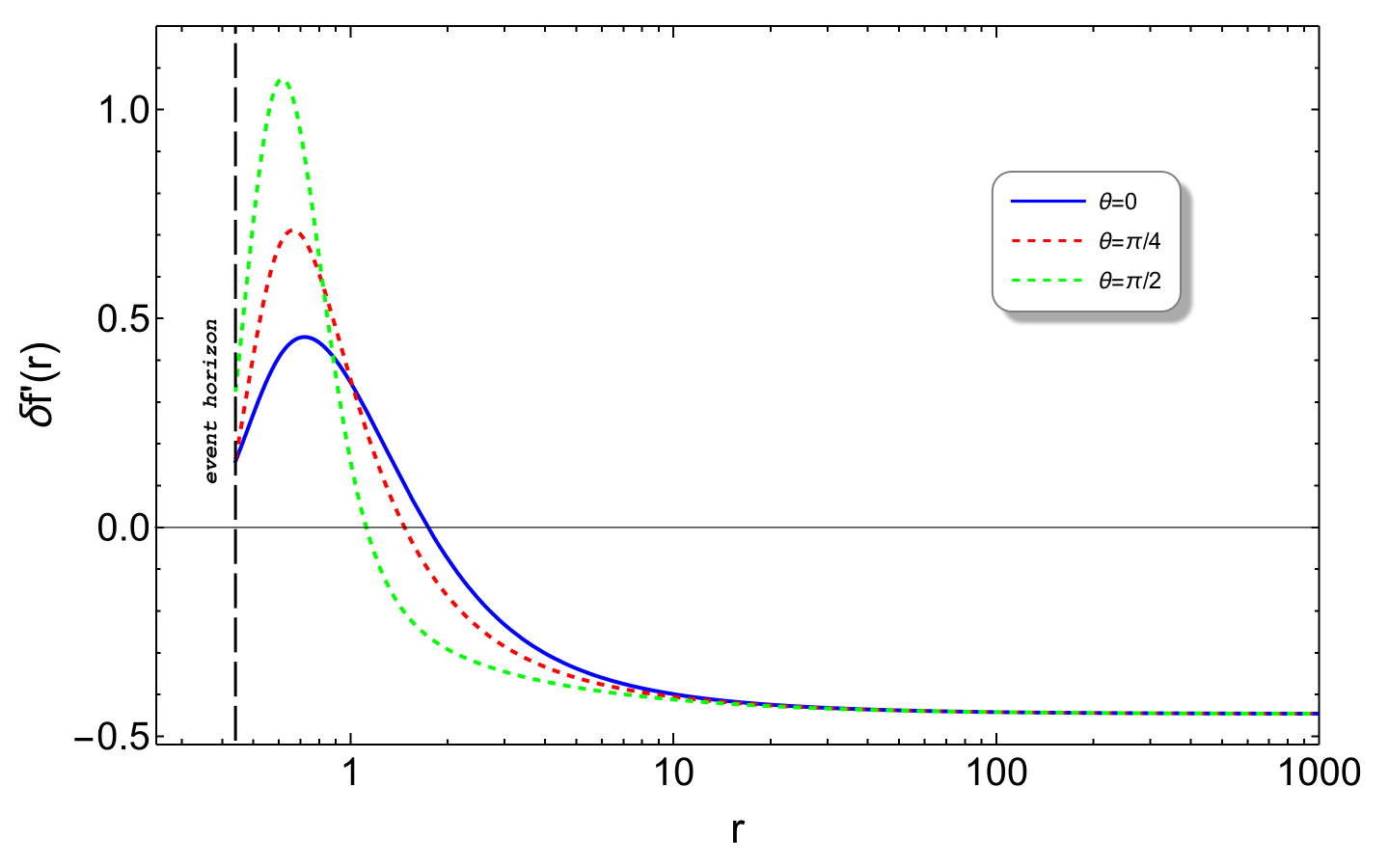}
    \end{minipage}
    \hspace{0.5cm}
    \begin{minipage}{0.4\textwidth}
	\centering
	\includegraphics[width=\textwidth]{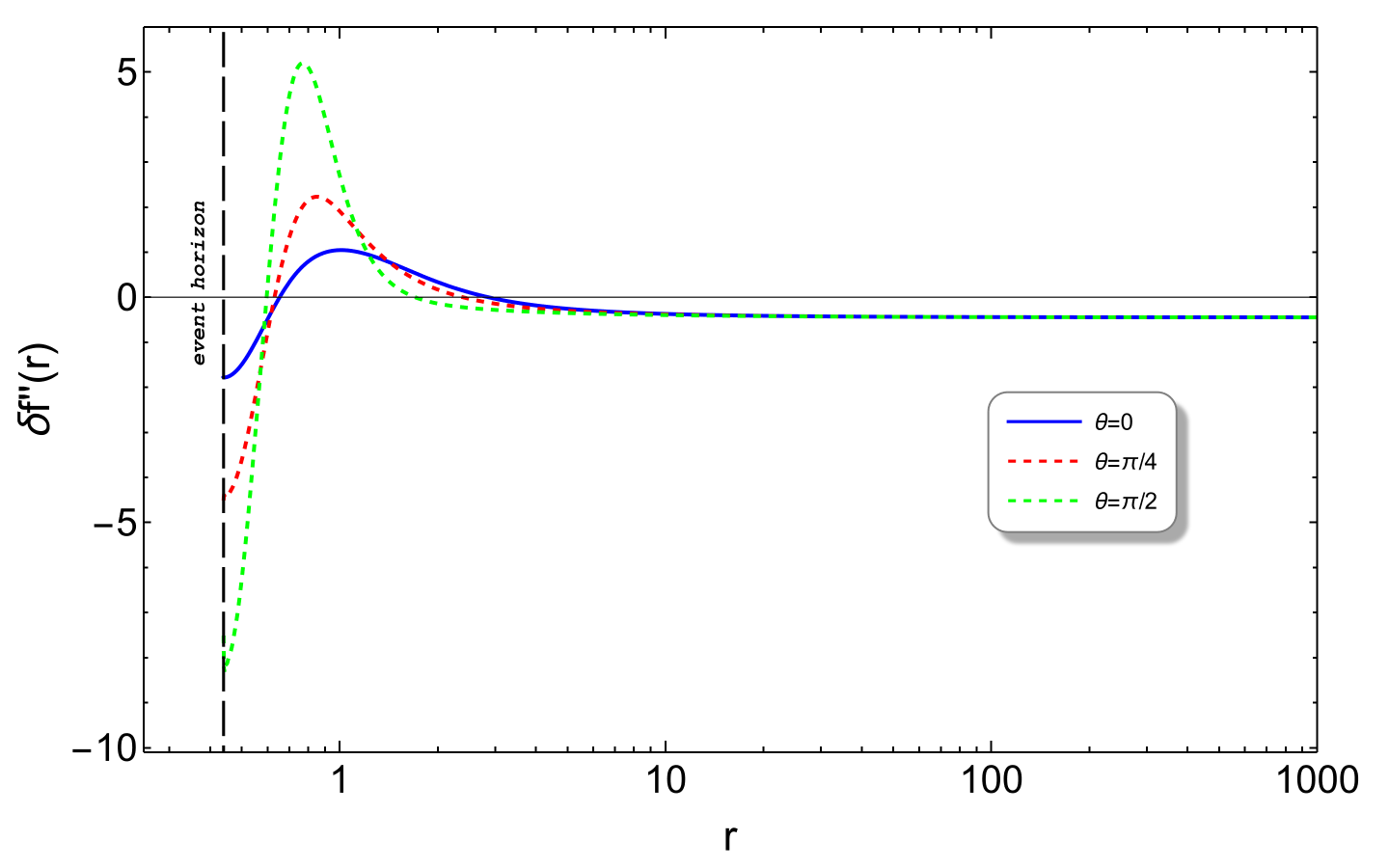}
    \end{minipage}
    \caption{The percentage change of the first order derivative (left panel) and second order derivative (right panel) of metric function $f$.}
    \label{fig:error df ddf}
\end{figure}

To assess the impact of the quartic term $\mathcal{G}^2$ on the spacetime geometry and scalar fields, we calculate numerical solutions for the model with $\alpha_2/\alpha_1=0.3$ and compare them with those obtained with $\alpha_2/\alpha_1=0$.
Using the same BH parameters $r_H=0.44$ and $\chi=0.5$ as in Figs. \ref{fig:3D 2D metric functions} and \ref{fig:3D 2D scalar fields}, we obtain an ADM mass $M=1.04453$ and a scalar charge of $Q_s=0.230125$.
To quantitatively describe the difference, we define the percentage change for each function as $\delta \mathcal{F}^{(k)}=(\mathcal{F}^{(k)}_{\alpha_{2}=0.3}-\mathcal{F}^{(k)}_{\alpha_{2}=0})/\mathcal{F}^{(k)}_{\alpha_{2}=0}\times 100$ with $\mathcal{F}^{(k)} =\{f,g,h,W,\phi,\psi\}$.
Fig. \ref{fig:error} shows the numerical results for $\delta \mathcal{F}^{(k)}$.
We observe that, except for $\psi$, the functions $\mathcal{F}^{(k)}$ exhibit only minor deviations from those in the case $\alpha_2/\alpha_1=0$.
The percentage change for the metric functions is of the order of $0.1\%$, while for the scalar field  $\phi$ , the maximum deviation is approximately $3\%$ near the horizon.
In contrast, the scalar field $\psi=\mathcal{G}$ shows a more pronounced difference, with deviations reaching around $20\%$ near the horizon.
This larger discrepancy arises because, although the scalar functions themselves differ only slightly, their second-order derivatives with respect to $r$ have significant fluctuations near the horizon.
These fluctuations amplify the variation in $\psi$, resulting in the observed difference.
For example, Fig.\ref{fig:error df ddf} gives the percentage change in the derivatives of the metric function $f$.

%%%%%%%%%%%%%%%%%%%%%%%%%%%
%%%%%%%%%%%%%%%%%%%%%%%%%%%
\section{Conclusions}
\label{Conclusions}
%%%%%%%%%%%%%%%%%%%%%%%%%%%
%%%%%%%%%%%%%%%%%%%%%%%%%%%
In this study, we explored how rotating BHs behave in the context of a class of sGB gravity theories that incorporate a squared Gauss-Bonnet term. We focus on three key factors: (1) the constant $\kappa$, which corresponds to the coupling strength between the gravitational field and scalar field; (2) the dimensionless spin parameter ($\chi$), which measures how fast a BH rotates, (3) and the coupling constant ($\alpha_2$), which represents additional gravitational effects in sGB gravity. Our goal is to understand how these factors influence the existence and properties of scalarized BHs. 
By numerically solving the field equations under stationary and axisymmetric conditions, we demonstrated that the constant $\kappa$ plays a critical role in shaping the characteristics of scalarized BHs in sGB gravity. Smaller $\kappa$ values enhance the coupling strength, leading to larger scalar charges.
Moreover, we demonstrated that the scalarization of BHs is significantly suppressed by the spin parameter ($\chi$) and the squared coupling constant ($\alpha_2$). Specifically, increasing $\alpha_2$ narrows the domain of existence for scalarized solutions, particularly for lower-mass BHs. This suppression arises from the $\mathcal{G}^2$-term in the effective mass $m^{2}_{eff}$, which stabilizes general relativity solutions against tachyonic instabilities.

Furthermore, scalarized BHs are shown to be thermodynamically favored over Kerr BHs with identical mass and spin in most parameter regimes, as evidenced by their higher entropy ($S/S_{GR}$ > 1). However, this entropic preference diminished for highly spinning BHs or large $\alpha_2$.
%, where the suppression effects dominate. 
These results challenge the "Kerr hypothesis" and highlight the role of modified gravity in enriching the BH phase space. The predicted scalarization mass windows could be probed by future gravitational wave detectors like LISA, Taiji and TianQin, offering observational constraints on the coupling constants $\alpha_1$ and $\alpha_2$.
It should be noted that the rotating scalarized BHs investigated in this work fall within the curvature-induced scalarization framework. In contrast, spin-induced scalarization, which represents a distinct mechanism and has garnered significant interest in recent studies \cite{PhysRevLett.125.231101,PhysRevD.102.084060,PhysRevD.102.124056,PhysRevD.102.104027}, remains an intriguing direction for future exploration.
Future work should focus on dynamical scenarios, such as BH mergers or accretion processes, to validate these predictions and refine the parameter space of sGB gravity.

%%%%%%%%%%%%%%%%%%%%%%%%%%%
%%%%%%%%%%%%%%%%%%%%%%%%%%%
\section*{ACKNOWLEDGMENTS}
%%%%%%%%%%%%%%%%%%%%%%%%%%%
%%%%%%%%%%%%%%%%%%%%%%%%%%%
This research is supported by the National Natural Science Foundation of China under Grant Nos.12375056, 12375048, and the Postgraduate Research $\&$ Practice Innovation Program of Jiangsu Province under Grant No.KYCX24\_3712.
Some of our calculations were performed using the tensor-algebra bundle xAct~\cite{xact}.

%%%%%%%%%%%%%%%%%%%%%%%%%%
%%%%%%%%%%%%%%%%%%%%%%%%%%
{\centering \section*{Appendix: Resolution Settings for Computation}}
\label{Appendix}
%%%%%%%%%%%%%%%%%%%%%%%%%%
%%%%%%%%%%%%%%%%%%%%%%%%%%

In this appendix, we provide an example to explain why we constructed a grid with $N_x$ = 40 and $N_{\theta}$ = 8 when using Chebyshev pseudo-spectral and Newton-Raphson methods to solve the system of field equations in this paper.

\begin{figure}[htbp]
    \centering
    \begin{minipage}{0.45\textwidth}
        \centering
	\includegraphics[width=\textwidth]{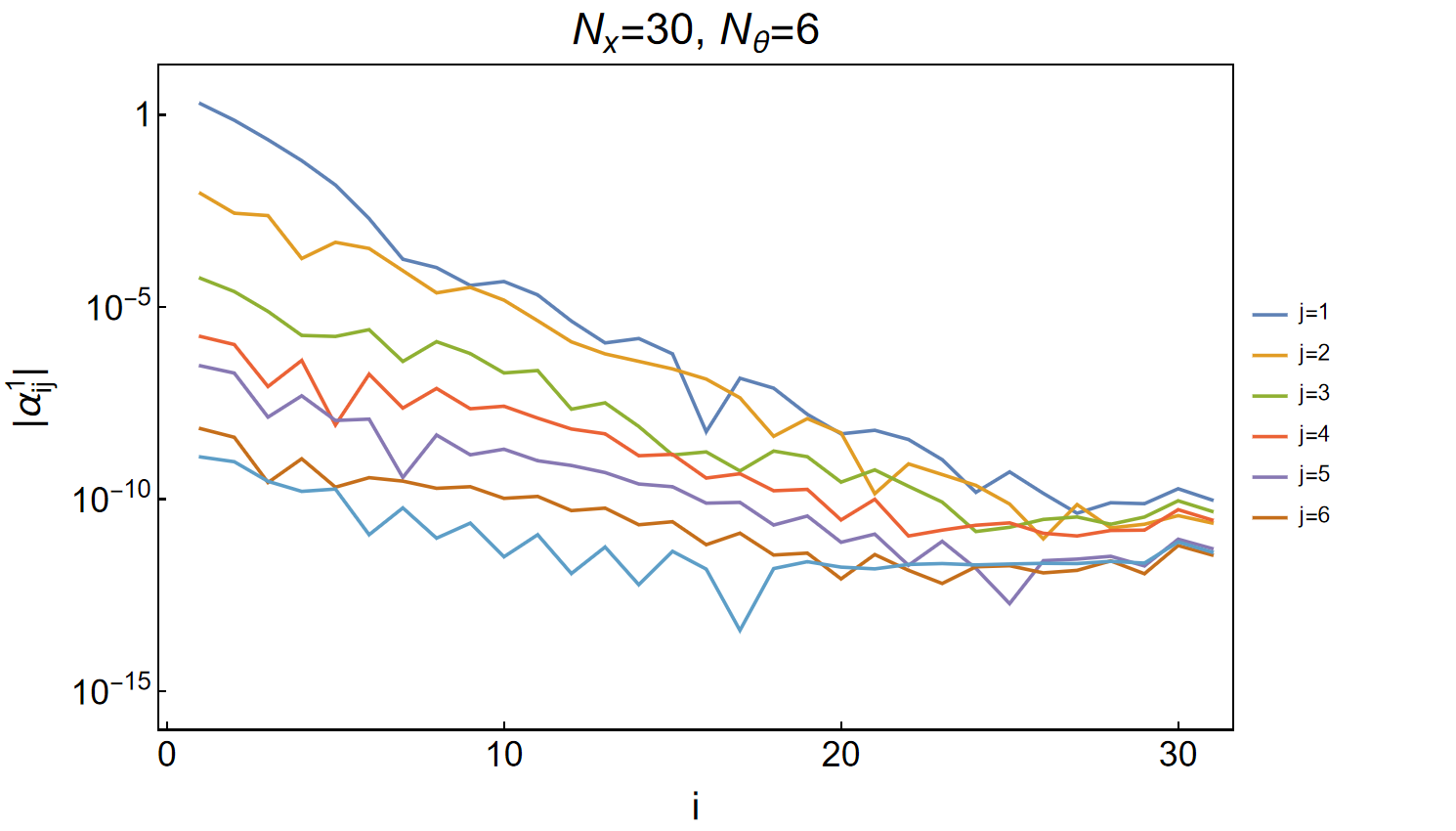}
    \end{minipage}
    \hspace{0.1cm}
    \begin{minipage}{0.45\textwidth}
	\centering
	\includegraphics[width=\textwidth]{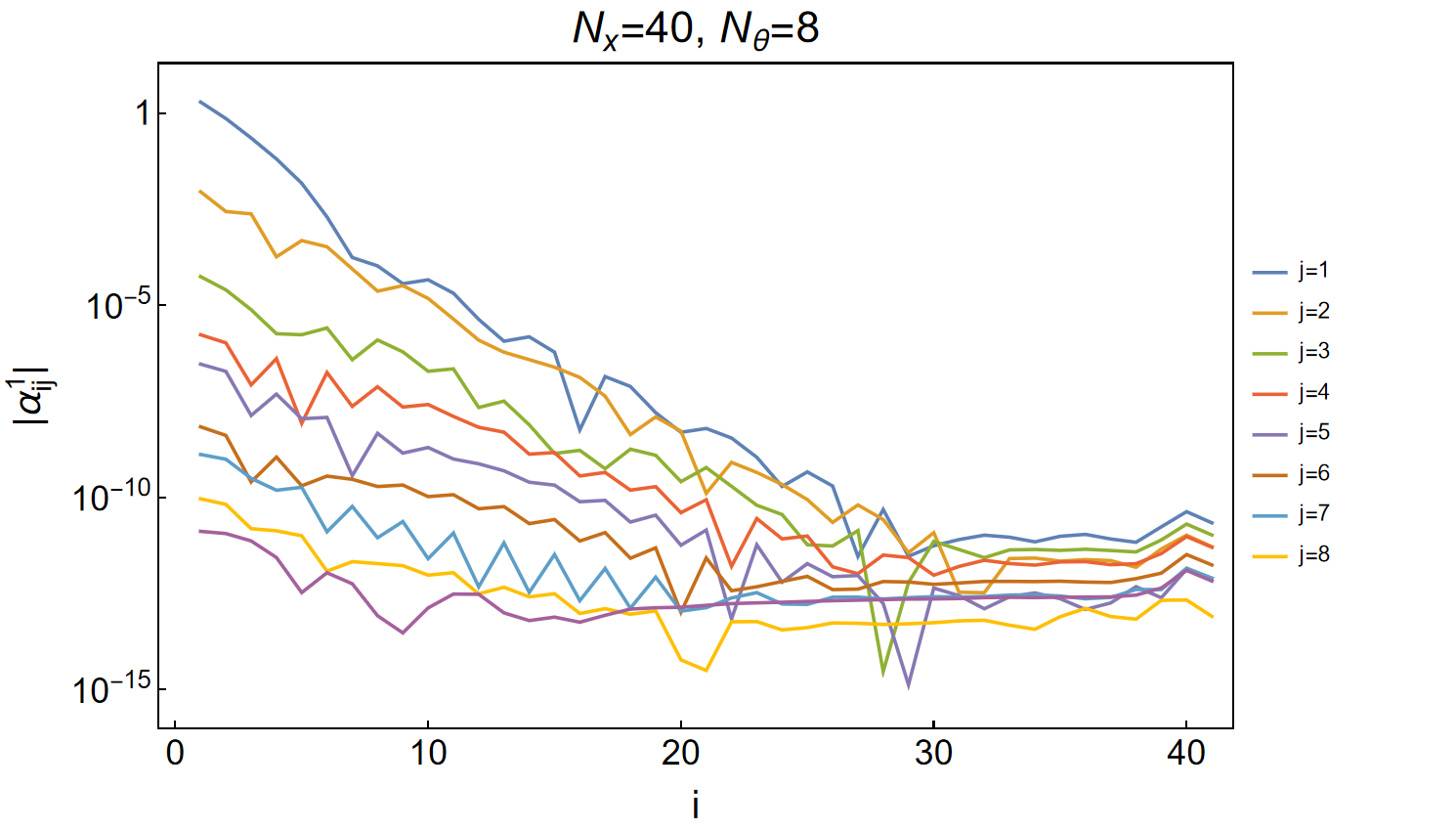}
    \end{minipage}
    
    \begin{minipage}{0.45\textwidth}
        \centering
	\includegraphics[width=\textwidth]{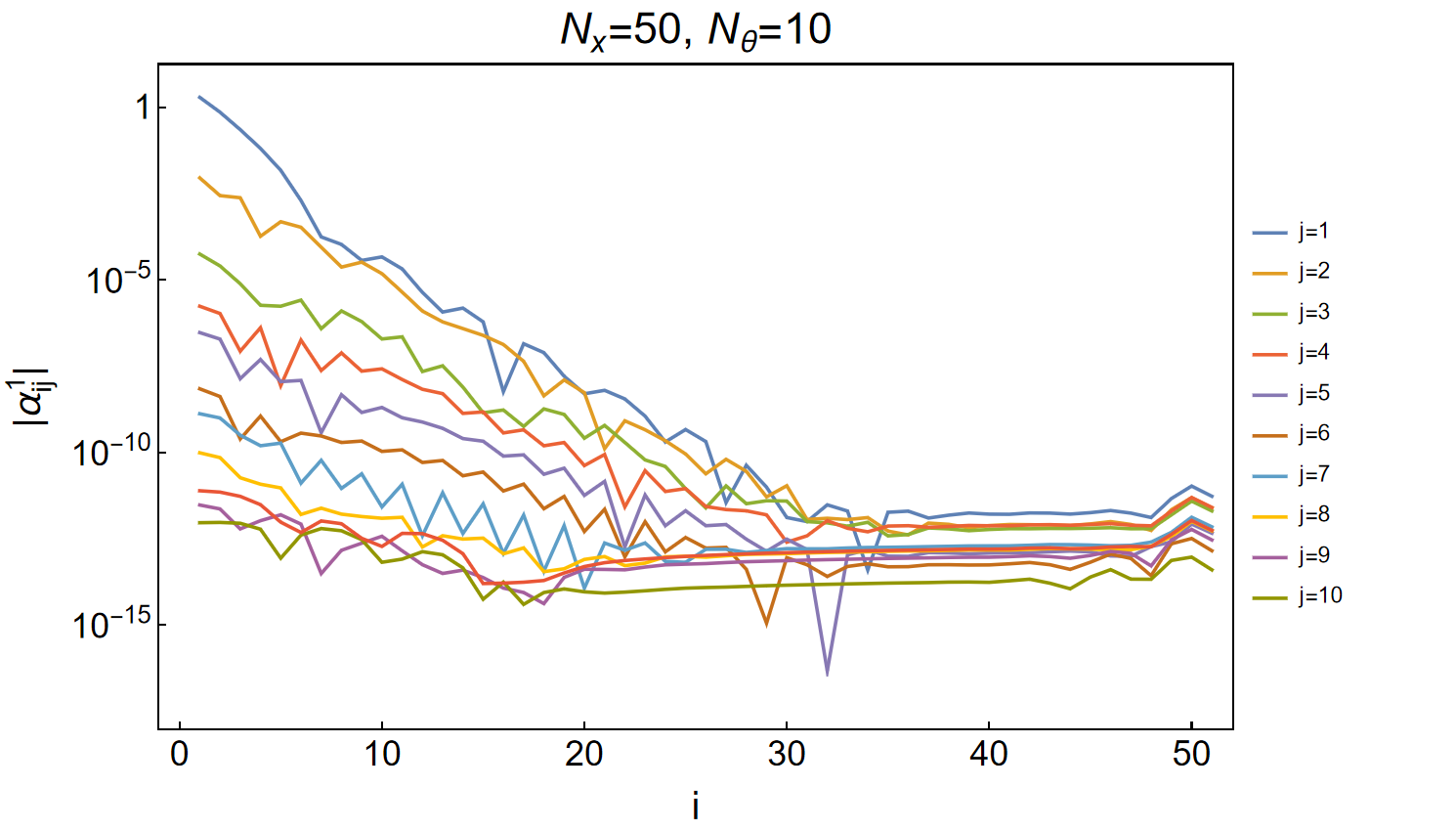}
    \end{minipage}
    \hspace{0.1cm}
    \begin{minipage}{0.45\textwidth}
	\centering
	\includegraphics[width=\textwidth]{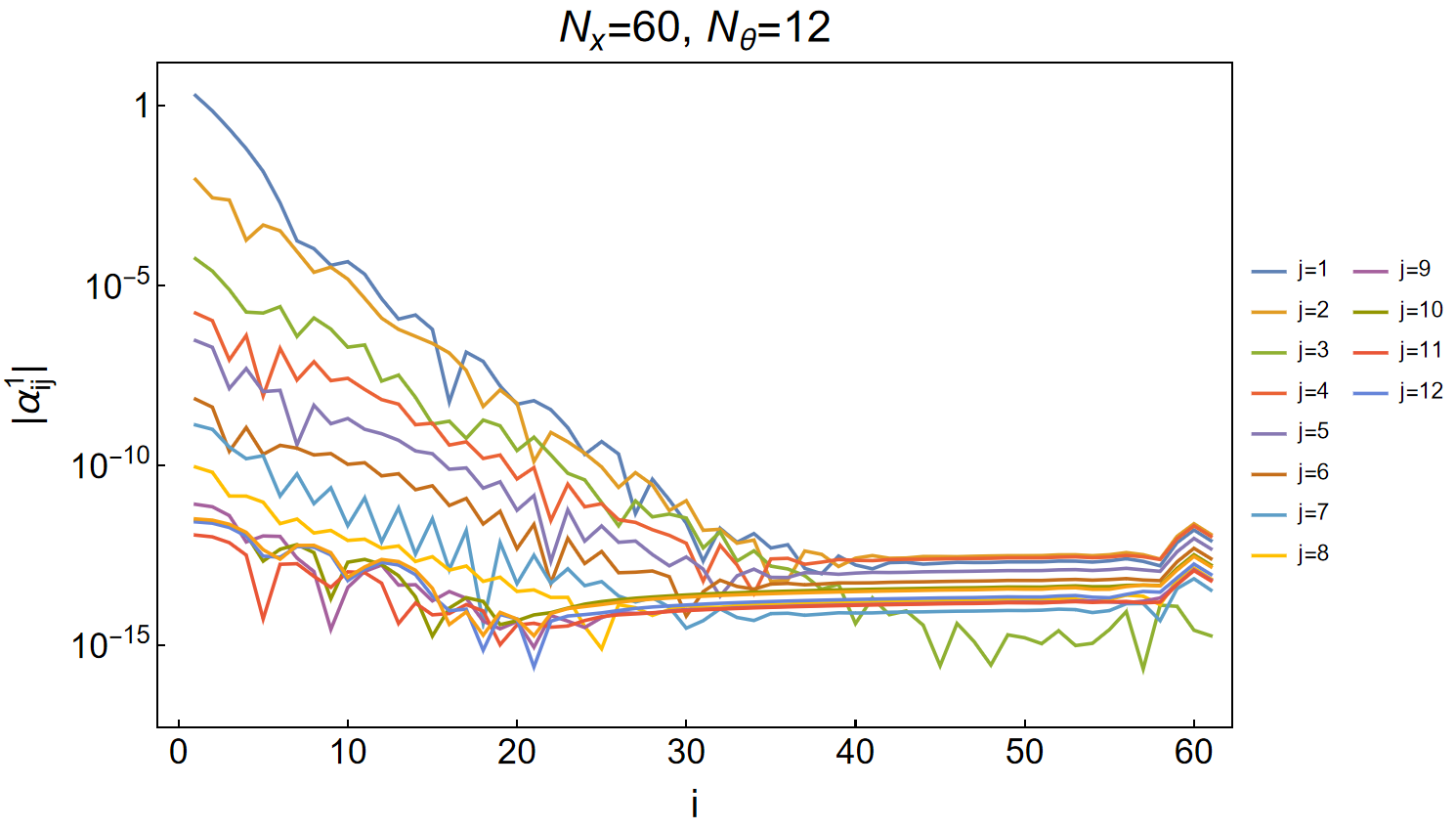}
    \end{minipage}
    
    \caption{
    The absolute values of the spectral decomposition coefficients $\alpha^{(1)}_{ij}$ of the metric function $f\left(x, \theta\right)$ with different $\left(N_x, N_{\theta}\right)$.
    }
    \label{fig:coefficient metric}
\end{figure}
Employing the dimensionless parameters $r_H$ = 0.46 (horizon radius) and dimensionless spin $\chi$ = 0.4, we present in Fig. \ref{fig:coefficient metric} the spectral decomposition coefficients $\alpha^{(1)}_{ij}$ (see Eq.\eqref{ExpandSerise}) governing the metric function $f\left(x, \theta\right)$. One can see from it that the magnitudes $|\alpha_{ij}^{1}|$ exhibit an exponential decay with increasing index values of i for each j, which demonstrates the convergent characteristic of numerical scheme.
Consequently, the functional value $f\left(x, \theta\right)$ predominantly depends on the leading terms in the $\alpha^{(1)}_{ij}$ series. 
Even when the values of $\left(N_x, N_{\theta}\right)$ increase from (40, 8) to (50, 10) and (60, 12), the original coefficients remain unchanged, and the additional coefficients become negligible. 
Moreover, if set $N_x$ = 30 and $N_{\theta}$ = 6, we are not sure that those leading terms could approximate the true scalarized BH solution. Through a comprehensive evaluation of computational efficiency and numerical accuracy, $N_x$ = 40 and $N_{\theta}$ = 8 have been identified as the optimal configurations for main computations in this work.

\centering
\bibliographystyle{unsrt}
\bibliography{BmyRef.bib}

\end{document}